# Ultraluminous X-Ray Sources

**S. N. Fabrika**[a, b, ]*, **K. E. Atapin**[c], **A. S. Vinokurov**[a], **and O. N. Sholukhova**[a]

[a] *Special Astrophysical Observatory, Russian Academy of Sciences, Nizhnii Arkhyz, 369167 Russia*
[b] *Kazan Federal University, Kazan, 420008 Russia*
[c] *Sternberg Astronomical Institute of Moscow State University, Moscow, 119992 Russia*
*e-mail: fabrika@sao.ru

**Abstract**—Ultraluminous X-ray sources (ULXs) were identified as a separate class of objects in 2000 based on data from the Chandra X-Ray Observatory. These are unique objects: their X-ray luminosities exceed the Eddington limit for a typical stellar-mass black hole. For a long time, the nature of ULXs remained unclear. However, the gradual accumulation of data, new results of X-ray and optical spectroscopy, and the study of the structure and energy of nebulae surrounding ULXs led to the understanding that most of the ultraluminous X-ray sources must be supercritical accretion disks like SS 433. The discovery of neutron stars in a number of objects only increased the confidence of the scientific community in the conclusions obtained, since the presence of neutron stars in such systems clearly indicates a supercritical accretion regime. In this review, we systematize the main facts about the observational manifestations of ULXs and SS 433 in the X-ray and optical ranges and discuss their explanation from the point of view of the supercritical accretion theory.

## 1. INTRODUCTION

The first evidence of the existence of abnormally luminous objects in the X-ray range in neighboring galaxies was obtained back in the 80's by the Einstein Observatory (Fabbiano, 1989). The luminosities of these objects were significantly higher than those typical for binary systems with black holes and neutron stars located in our Galaxy, but at the same time, the high spatial resolution of the X-ray telescope allowed us to state with confidence that these objects cannot be active galactic nuclei as well. With the launch of the Chandra X-Ray Observatory in 1999, it became clear that these "anomalous" sources are not isolated finds, but represent a new, rather extensive class of objects, later called *ultraluminous X-ray sources, ULXs*.

So, ultraluminous X-ray sources are point-like extragalactic objects with a luminosity above $2 \times 10^{39}$ erg s$^{-1}$ (the Eddington limit for a black hole of mass 15 $M_{\odot}$ of the solar chemical composition, $L_{\text{Edd}} \approx 1.5 \times 10^{38} m$ erg s$^{-1}$, where $m$ is the mass in units of Solar mass), but are not active galactic nuclei (Kaaret et al., 2017). In total, several hundred such objects are currently known (Earnshaw et al., 2019; Liu, 2011; Swartz et al., 2004; Walton et al., 2011). The vast majority of them are located in spiral and irregular galaxies with a high rate of star forming and are associated with the youngest stellar population. In particular, using the example of the Antenna galaxy, it was shown that all the brightest

X-ray sources were formed in the centers of very young clusters with an age of less than 5 Myr, and then, as a result of the dynamic evolution of clusters, they were ejected from them, flying out at distances up to 200 pc (Poutanen et al., 2013).

Initially, two main models were considered as an explanation for the ULX phenomenon. The first of them claimed that these systems can contain so-called *intermediate mass black holes (IMBHs)* with masses of $10^3-10^5$ $M_{\odot}$ (Colbert et al., 1999), which occupy an intermediate position between stellar-mass black holes in binary systems and supermassive black holes in the galactic nuclei. In the case of accretion on IMBH, it is not necessary to exceed the Eddington limit to obtain the observed luminosities, so it was expected that such objects should accret in the same (subcritical) mode as the X-ray sources of our Galaxy, and have generally similar observational manifestations. An alternative model, now the dominant one, assumed that ULXs are close binary systems with stellar-mass black holes (or neutron stars), in which the donor star fills the Roche lobe and accretion occurs in the super-Eddington (supercritical) mode (Fabrika and Mescheryakov, 2001; King et al., 2001). In this case, a supercritical accretion disk should be formed (Lipunova, 1999; Poutanen et al., 2007; Shakura and Sunyaev, 1973), which differs significantly from the standard one. Some unusual ideas were also expressed, for example, a model was proposed in which the source of ULX energy is not accretion, but the rotation energy of the newly born pulsar (Medvedev and Poutanen, 2013).

The main distinguishing feature of a supercritical disk is the presence of powerful outflows of matter (or wind) from its center; evidence of such outflows was found several years ago in a number of objects in both the optical and X-ray ranges (see below). Another significant result of recent years is the detection of coherent pulsations in the object M 82 X-2 (Bachetti et al., 2014), which made it the first representative of a new class of sources—*ultraluminous X-ray pulsars (ULXP)*. The presence of pulsations means that the accretor must be a neutron star, that is, the Eddington limit in these objects can be exceeded several hundred times (under the condition of isotropic radiation). Thus, the discoveries of the last few years suggest that most of the ULXs appear to be superaccretors. However, some of the objects (Sutton et al., 2012), including the hyperluminous X-ray source ESO 243-49 HLX-1 (Davis et al., 2011; Farrell et al., 2009; Titarchuk et al., 2016) (luminosity reaches $10^{42}$ erg s$^{-1}$), can still be systems with IMBH.

There is also one known object in our Galaxy that accretes in the super-Eddington mode—this is the SS 433 system (Fabrika, 2004). This object is famous for its constant jets of gas ejected from the center of the accretion disk at a speed of $V_j \approx 0.26c$. The kinetic energy of the jets is estimated at about $10^{39}$ erg s$^{-1}$ (Panferov and Fabrika, 1997), and the total energy of the system at $L_{bol} \sim 10^{40}$ erg s$^{-1}$, which corresponds to the luminosity of bright ULXs. However, the orientation of SS 433 is such that the inner parts of its supercritical accretion disk are always shielded from the observer by an optically thick wind and the maximum of its radiation falls on the optical/UV ranges. Therefore, in X-rays, SS 433 is a relatively weak source with a luminosity of $L_X \sim 10^{36}$ ergs$^{-1}$, and the main source of radiation reaching the observer are the hot parts of the jets (Medvedev and Fabrika, 2010). The accretion rate in the SS 433 system is estimated at $\dot{M}_0 \approx 300 M_\odot$ yr$^{-1}$ (Fabrika, 2004).

In 2001, we published an article Fabrika and Mescheryakov (2001), in which we suggested that in other galaxies there should be objects similar to SS 433, but viewed face-on and being bright X-ray sources. Also, using data from the ROSAT catalog of bright sources (18 811 sources) and the RC3 catalog (about 19 000 galaxies), we estimated the frequency of occurrence of such objects: for spiral and irregular galaxies at distances up to 10 Mpc we obtained frequencies 4–5%.

The concept of supercritical accretion disks was proposed by Shakura and Sunyaev (1973) and further developed in Abramowicz et al. (1988); Jaroszynski et al. (1980); Lipunova (1999); Poutanen et al. (2007). If the rate of matter supply at the outer edge of the disk $\dot{M}_0$ exceeds the critical level (corresponding to the Eddington luminosity) $\dot{M}_{Edd} = 48\pi GM/c\kappa \approx 2 \times 10^{18} m$ g s$^{-1}$, where $\kappa$ is the Thomson opacity (Poutanen et al., 2007), then in the inner part of the disk there is a region in which the radiation pressure force becomes equal to the force of gravity. The size of this area, bounded by *spherization radius*, is directly proportional to the rate of accretion $r_{sp} \approx (5/3)\dot{m}_0$ (Poutanen et al., 2007), where $\dot{m}_0 = \dot{M}_0/\dot{M}_{Edd} \gg 1$ is the dimensionless accretion rate, and the dimensionless spherization radius $r_{sp}$ is expressed in units of the inner radius of the disk $R_{in}$, usually assumed to be equal to three Schwarzschild radii: $R_g = 2GM/c^2$ (Fig. 1).

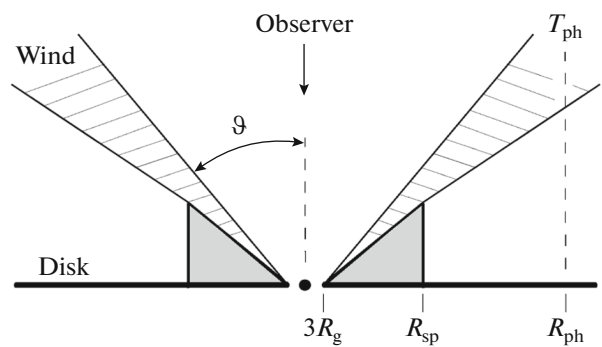

**Fig. 1.** Diagram of a supercritical accretion disk. Three radii are designated: the inner disk radius $R_{in} = 3R_g$, the spherization radius $R_{sp}$ and the radius of the wind photosphere $R_{ph}$. The supercritical properties of the disk appear below the spherization radius, where the disk becomes geometrically thick $H(R) \simeq R$. In the area $R \gtrsim R_{sp}$ the disk is similar to the standard (thin) one. It is assumed that the wind forms a cone-shaped funnel with an opening angle $\vartheta \sim 20°$–$50°$ and is practically not filled with matter inside. In the case of the ULX, the observer must "look" directly into the funnel and see the hot inner parts of the disk emitting in the X-ray range.

Supercritical properties of the disk appear inside the spherization radius: it becomes geometrically thick with $H/R \sim 1$, where $R$ is the distance to the black hole, and $H$ is the half-thickness of the disk, and there is an outflow of matter in the form of a powerful wind. The wind must have an angular momentum, so it is assumed that wind forms a funnel, having the shape of a hollow cone (Fig. 1). Due to the outflow of matter, as we approach the black hole, the amount of matter $\dot{M}(R)$ passing through the disk decreases, and eventually only a relatively small part of the matter reaches the inner boundary of the disk. Advection is important here—the process of carrying away energy together with matter into a black hole due to photon trapping, since it reduces the radiation pressure on the disk surface and weakens the wind (Lipunova, 1999; Poutanen et al., 2007).

The bolometric luminosity of a supercritical disk exceeds the Eddington luminosity by a logarithmic factor $L_{\mathrm{bol}} \approx L_{\mathrm{Edd}}\left(1 + \frac{5}{3}\ln\dot{m}_0\right)$ (Poutanen et al., 2007), i.e. at the accretion rate of $\dot{m}_0 \sim 100$, the Eddington luminosity will be exceeded by about 4 times. In addition, the observed luminosity can increase by an additional 2–7 times due to the radiation beaming in the funnel (Fabrika and Mescheryakov, 2001; King et al., 2001). The hottest part of the disk is located near its inner boundary and has a temperature of $T_{\mathrm{max}} \approx 1.6 m^{-1/4}$ keV $\simeq 1$ keV (for the accretor mass 10 $M_\odot$). The temperature at the spherization radius depends on the accretion rate $T_{\mathrm{sp}} \approx \dot{m}_0^{-1/2} T_{\mathrm{max}}$ and for $\dot{m}_0 \sim 100$ is approximately 0.1 keV (Poutanen et al., 2007).

The rate of matter outflow from the supercritical disk is $\dot{M}_w = a\dot{M}_0$, where $a \lesssim 0.58$ is the parameter that depends on how much of the energy released in the disk is spent on wind acceleration (Poutanen et al., 2007). Thus, more than half of the matter entering the disk from the donor star can be ejected as wind. The wind forms a photosphere whose radius $r_{\mathrm{ph}} \propto \dot{m}_0^{3/2}$ (in units of $R_{\mathrm{in}}$) can be several orders of magnitude greater than the spherization radius. The temperature at the photosphere radius decreases with increasing accretion rate $T_{\mathrm{ph}} \propto \dot{m}_0^{-3/4}$ (Poutanen et al., 2007), and the maximum photospheric radiation is always in the ultraviolet range. The presence of powerful ultraviolet radiation is confirmed by photoionization nebulae surrounding many ultraluminous X-ray sources (see, for example, Abolmasov et al. (2007, 2008); Kaaret et al. (2010)).

The above concepts of supercritical accretion are in good agreement with both the MHD calculations (Okuda et al., 2009; Ohsuga and Mineshige, 2011; Ohsuga et al., 2005; Takahashi et al., 2018; Takeuchi et al., 2013, 2014) and the observations of real objects: ULXs and SS 433. Below we will consider in detail the observational manifestations of ultraluminous X-ray sources, simultaneously comparing them with the object SS 433. In Section 2, the X-ray radiation of ULXs will be described: X-ray spectra, variability, and properties of ultraluminous pulsars. In Section 3 we will talk about optical radiation. Optical spectroscopy and photometry provide a lot of useful information not only about the ULXs themselves, but also about their surroundings, which can shed light on their origin and evolution as binary systems.

## 2. X-RAY RADIATION FROM ULXS

### 2.1. X-Ray Spectra

When the XMM-Newton and Chandra space telescopes obtained observations of ULXs with a sufficiently high accumulation, it became clear that the ULX spectra are significantly different (Gladstone et al., 2009; Stobbart et al., 2006) from those observed in the X-ray binaries of our Galaxy (McClintock and Remillard, 2006). Most of the spectra have a double-peaked shape with an inflection in the region of 2 keV and an exponential cutoff above 10 keV (Bachetti et al., 2013; Walton et al., 2014). The formal approximation of such spectra by the two-component model "standard disk + comptonization," which is widely applied to the spectra of Galactic X-ray binaries, gives the temperatures of the disk and the comptonizing medium much lower, and the optical depth higher: $kT_{\mathrm{disk}} \sim 0.2$ keV, $kT_e \sim 1-2$ keV and $\tau \gtrsim 6$ for ULXs (Gladstone et al., 2009) versus $kT_{\mathrm{disk}} \sim 1$ keV, $kT_e \sim 100$ keV and $\tau \lesssim 1$ for X-ray binaries (McClintock and Remillard, 2006). The state with such an atypical spectral shape for objects in our Galaxy was proposed to be called "ultraluminous" state (Gladstone et al., 2009).

Depending on which of the two components of the spectrum contributes more, the ULX spectra can be divided into soft (soft ultraluminous, SUL) and hard (hard ultraluminous, HUL) ones (Sutton et al., 2013). In addition, the "broadened disk" (BD) type is distinguished, which got its name because the best agreement with the observations is given by the pure disk model with a temperature of $1-2.5$ keV. In spectra of this type, there is no obvious separation into humps and there is a maximum in the region of 5 keV. Examples of spectra of various types are shown in Fig. 2.

The spectrum of a particular object can change, passing from one state to another; nevertheless, there is a tendency for separation of the sources into soft and hard ones. When analyzing a sample of two dozen objects, it turned out that BD-type spectra are characteristic, rather, of weaker ULXs (Sutton et al., 2013, 2017), however, if a source showing a two-component spectrum (HUL or SUL) most of the time becomes brighter, then its spectrum also becomes similar to the BD type (Luangtip et al., 2016; Pintore and Zampieri, 2012; Pintore et al., 2014; Shidatsu et al., 2017). Fig. 3 shows the spectra of the object IC 342 X-1 in different states. In the low state (the luminosity in the range $0.3-30$ keV is $L_X \simeq 1.5 \times 10^{40}$ erg s$^{-1}$) the spectrum has an explicit power-law section and can be classified as the HUL type. In the bright state ($L_X \simeq 3.4 \times 10^{40}$ erg s$^{-1}$) the spectrum is convex (BD-type). It is interesting that during a long observation lasting about $10^5$ s, the source weakened three times, while the spectrum became more hard.

**Ultraluminous supersoft sources (ULS)** seem to be another type of ULXs (Liu and Di Stefano, 2008; Soria and Kong, 2016; Urquhart and Soria, 2016a). These objects are characterized by very soft thermal spectra with temperatures below 0.1 keV and almost no radiation above 1 keV. Using the example of seven objects, it was shown that the size of the thermal radiation source (photosphere) correlates with the temperature, but the bolometric luminosity remains con-stant (Urquhart and Soria, 2016a). When the temperature drops below 50 eV, all the radiation from the source moves into the ultraviolet range, and it becomes inaccessible to X-ray telescopes. When the temperature rises above 150 eV, a ULS turns into a ULX with a soft spectrum type (Feng et al., 2016; Pinto et al., 2017a; Urquhart and Soria, 2016a).

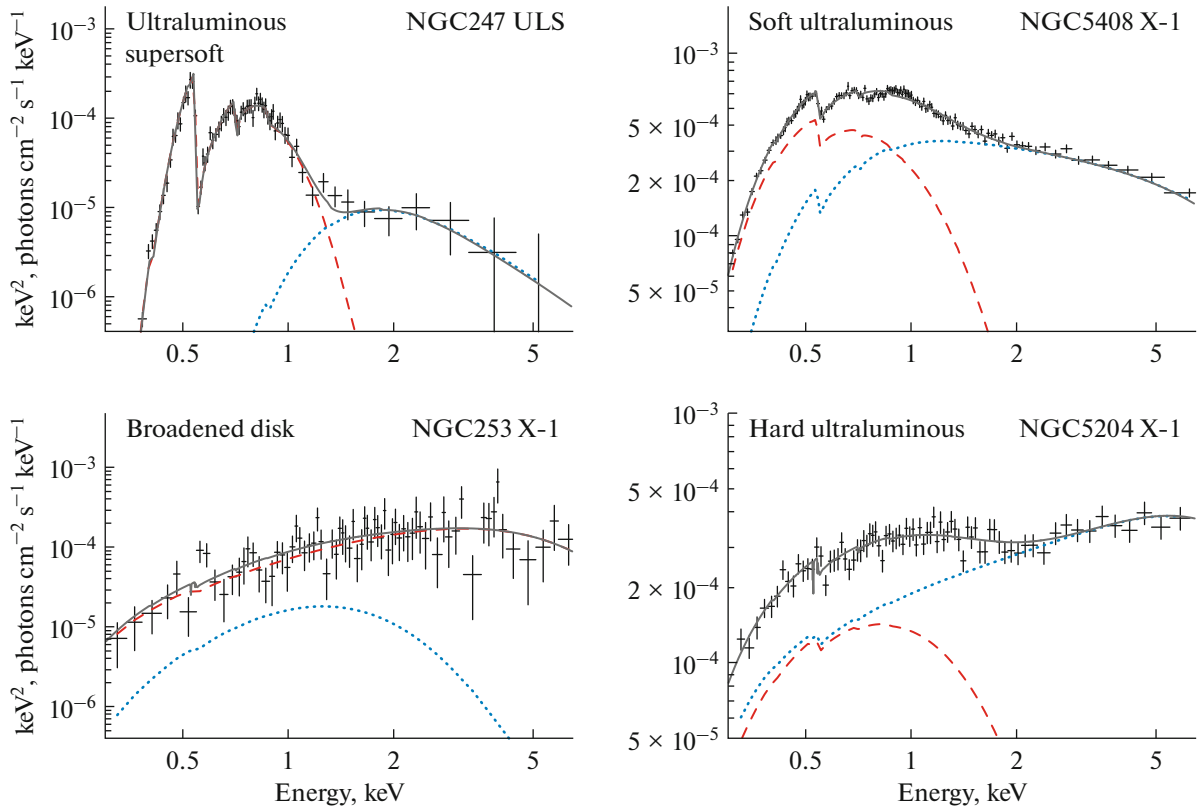

**Fig. 2.** Four types of ultraluminous X-ray source spectra in order of increasing hardness. (a) The spectrum of the ultraluminous supersoft source NGC247 ULS. (b) The spectrum type "ultraluminous soft" by the example of the source NGC5408 X-1. (c) The type "broadened disk", NGC253 X-1. (d) "Ultraluminous hard", the object NGC5204 X-1. Each spectrum is approximated by a model (gray solid line) consisting of two components: a multicolor disk (dashed red line) and comptonization (dashed blue line).

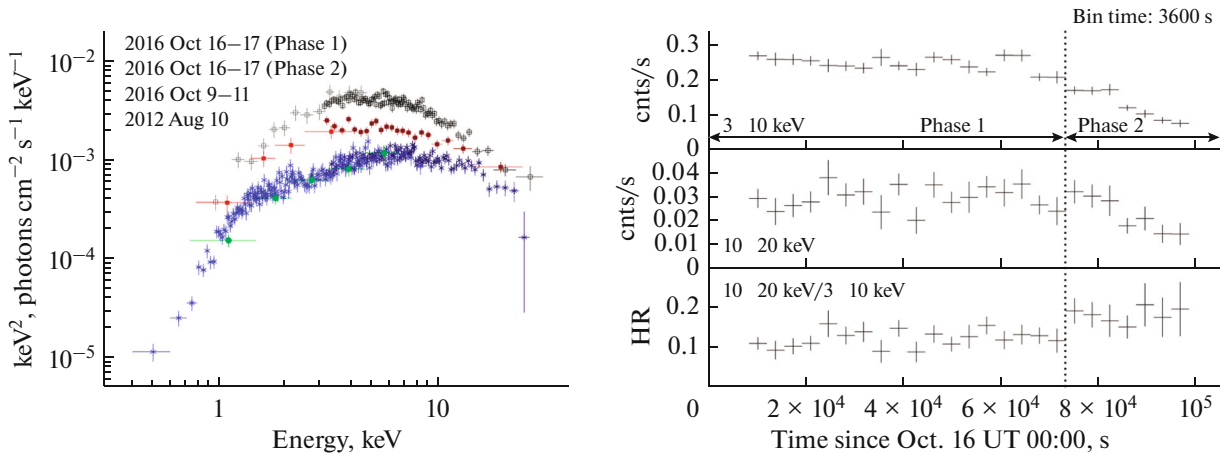

**Fig. 3.** Spectra (a) and light curves (b) of IC 342 X-1 from the work Shidatsu et al. (2017) based on monitoring data collected using the Swift, NuStar, and XMM-Newton X-ray telescopes. The light curves were obtained in two X-ray sub-bands on October 17, 2016 using the NuStar/FPMA instrument (their hardness ratio is also shown). During this observation, there was a noticeable decrease in the brightness of the object, so the data was divided into two phases. The spectra (combined Swift + NuStar) of the first and second phases are shown as black and red dots, respectively. Green shows the spectrum obtained from Swift observations made a few days earlier. The blue color shows the spectrum from simultaneous XMM-Newton + NuStar observations made in 2012. It can be seen that in the bright state the spectrum has a more convex shape.

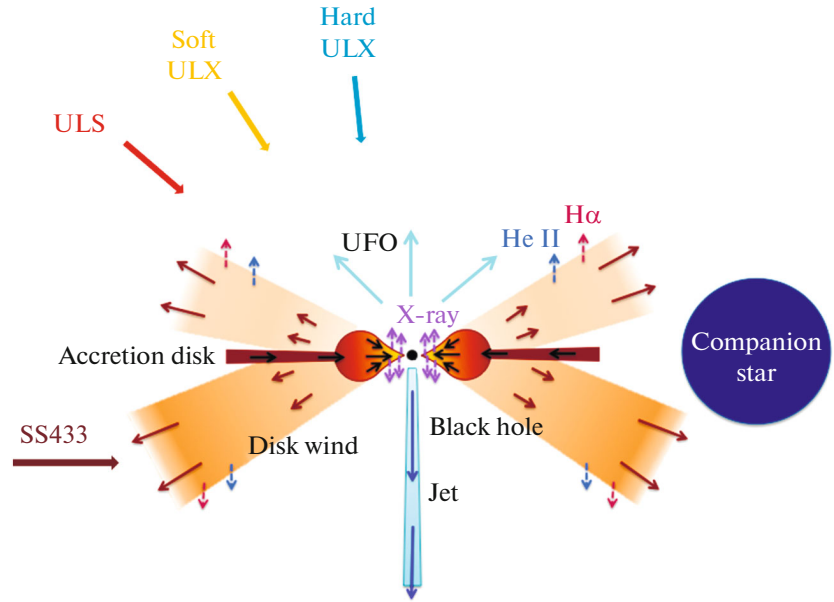

**Fig. 4.** The relationship between the type of object's spectrum and the angle at which the supercritical disk is observed. As the angle between the line of sight and the funnel axis increases, the spectrum should become softer. In the case of ULXs and ULSs, the observer looks into the funnel and sees the X-ray radiation of its internal parts. In the SS 433 system, the supercritical disk is visible from the side, its observed X-ray luminosity is less than 0.1% of the real one. The peripheral parts of the wind are a source of optical radiation. In addition to SS 433, relativistic jets were explicitly found only in one object—M 81 ULS (Liu et al., 2015), and in Holmberg II X-1, evidence of the presence of jets in the past was found (Cseh et al., 2014). However, many objects exhibit optically thin ultrafast outflows (UFOs), which appear to be an uncollimated jet (see text).

Despite the absence of physically motivated spectral models that could be directly fit all the available observations, in general, the variety of ULX spectral types is consistent with ideas of supercritical accretion. MHD calculations show that the conical funnel formed by the wind (Fig. 1) should have an opening angle of $\vartheta \sim 20°-50°$ (Okuda et al., 2009; Ohsuga et al., 2005; Takeuchi et al., 2013). Obviously, in the case of ULXs, the observer must directly "look" into the funnel and see the hottest parts of the supercritical disk. However, the angle $\theta$ at which he sees different objects (the angle between the funnel axis and the line of sight, Fig. 4), can be different (Pinto et al., 2017a; Pintore et al., 2014; Sutton et al., 2013). Objects observed almost along the funnel axis must have the most hard spectrum (BD and HUL objects, $\theta \simeq 0$). As the $\theta$ angle increases, the spectrum should become softer (SUL objects, $\theta < \vartheta$) due to the fact that the deepest parts of the funnel become less visible, as well as due to the fact that the line of sight begins to cross the wind clumps that break out of the funnel walls. With a further increase in the $\theta$ angle, the optical thickness of the wind increases, and wind clumps can completely cover the hard X-ray radiation—such objects look like ULSs ($\theta \approx \vartheta$). Finally, when $\theta \gg \vartheta$, objects of the SS 433 type should be observed, from which the X-ray radiation of the funnel reaches the observer only in the reflected form (Medvedev and Fabrika, 2010).

Both angles—both $\theta$ and $\vartheta$—can change over time, which causes the spectral variability of objects (Middleton et al., 2015a; Pinto et al., 2017a; Pintore et al., 2014). The $\theta$ angle can change due to the precession of the accretion disk (Dauser et al., 2017; Weng et al., 2018), which is well known in the example of SS 433 (Fabrika, 2004). The funnel opening $\vartheta$ probably depends on the accretion rate: as $\dot{m}_0$ increases, the wind increases and the funnel becomes narrower (King, 2009).

Based on the idea of a supercritical accretion disk of SS 433 and assuming that the accretor is a black hole, we developed a simplified model of a multicolor funnel (MCF, similar to a multicolor disk, MCD), which could predict what the spectra of similar objects should look like being observed face-on (Fabrika et al., 2006). To describe the funnel wall temperature distribution, we considered two limiting cases: one was dominated by the gas pressure $T(r) \propto r^{-1}$, and the other—by radiation pressure $T(r) \propto r^{-1/2}$. The temperature of the outer photosphere of the funnel and its depth are known from observations (the level of the photosphere at the bottom of the funnel was calculated based on the known mass flow rate in the jets). The temperature of the funnel walls in the deepest parts that can still be observed is estimated at $1 \times 10^6 – 1.7 \times 10^7$ K.

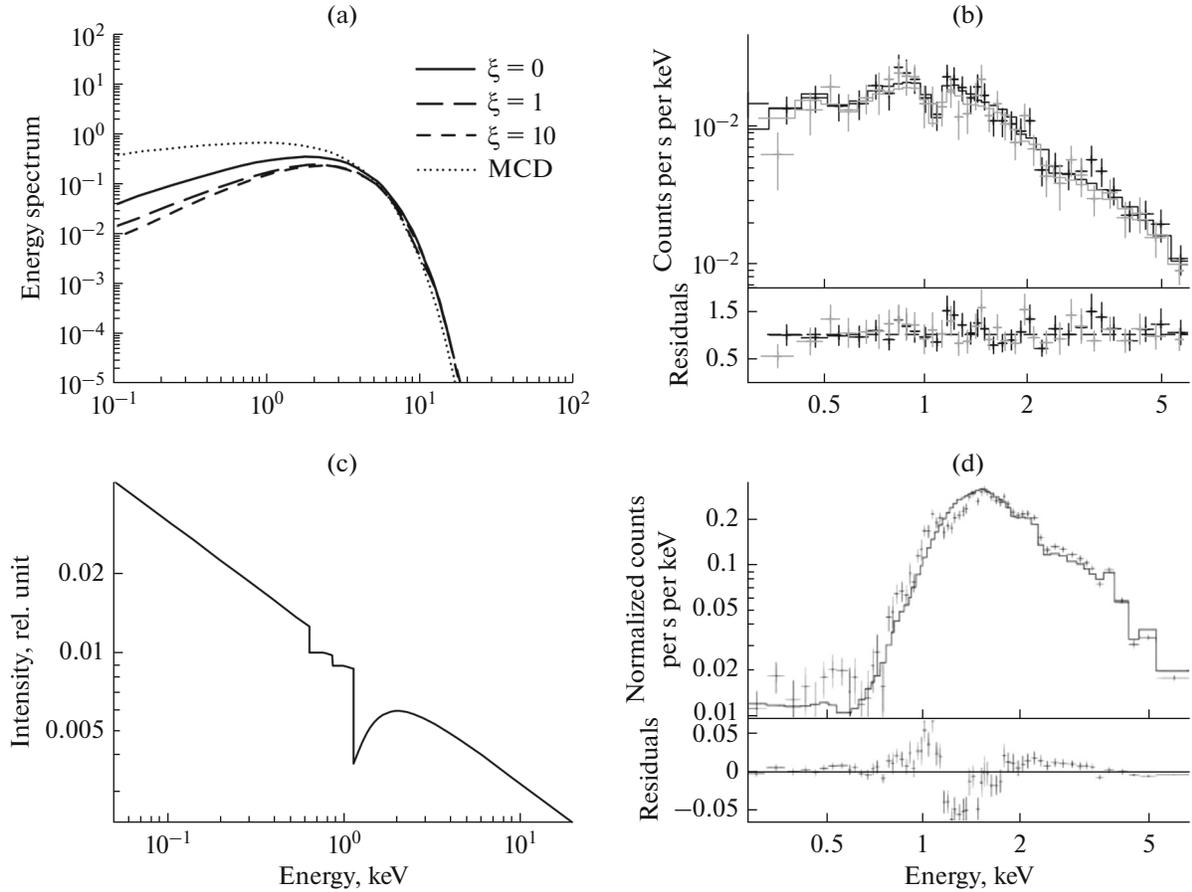

**Fig. 5.** (a) Model of continuous radiation of a multicolor funnel (MCF) in comparison with the model of a multicolor (standard) disk (MCD), ζ is the ratio of radiation pressure to gas pressure. (b) Application of the continuum funnel model to the observed spectrum of the ultraluminous X-ray source NGC 4736 X-1. (c) Model of emission/absorption features that can be observed in ULXs and in SS 433. A power spectrum with the exponent $\Gamma = 2.5$ is given with superimposed $L_c$-edges of ions C VI (about 0.63 keV), N VII (about 0.9 keV) and O VIII (about 1.1 keV) blue-shifted by $0.26c$. (d) It is shown how this spectrum would look on the XMM-Newton/MOS instrument with deep accumulation (more than $10^5$ photons), and also its approximation by a power law is given. In the lower part of the figure, residuals are shown, which should have a specific shape with a rise in the region of 1 keV and a dip immediately following it, repeating the shape of the O VIII edge (Fabrika et al., 2007).

In Fig. 5a we present the model MCF spectra for various values of the parameter $\zeta = aT_0^3/3k_b n_0$ —the ratio of radiation pressure to gas pressure; here $a = 4\sigma/c$, where σ is the Stefan–Boltzmann constant, $k_b$ is the Boltzmann constant, and $n_0$ is the density in the deepest parts of the funnel $r_0$. We took the temperature $T_0(r_0)$ equal to 1 keV. For comparison, the same figure shows a model of a multicolor disk (MCD) with $T_{in} = 1$ keV, as well as the MCF model applied to the observed spectrum of the ultraluminous X-ray source NGC 4736 X-1 (Fig. 5b).

The observed velocity of the jets of SS 433 $V_j \approx 0.26c$ indicates that the line-locking mechanism, well known from quasars (Shapiro et al., 1986), plays an important role in the acceleration of the jets. This means that the spectrum emitted by the funnel wall matter must be in the absorption containing $Lc$ and $Kc$ edges of hydrogen and helium-like ions (Fig. 5c). The resulting spectrum predicted by the model has a very complex structure consisting of a blend $K\alpha/Kc$ and $L\alpha/Lc$ of the most abundant heavy elements. In addition, its shape will be affected by changes in the physical parameters of the gas in the funnel: velocity, density, temperature, filling factor, etc. When such spectra are approximated by smooth continuum models (for example, by a power law), absorption edges should appear in residuals (Fig. 5d).

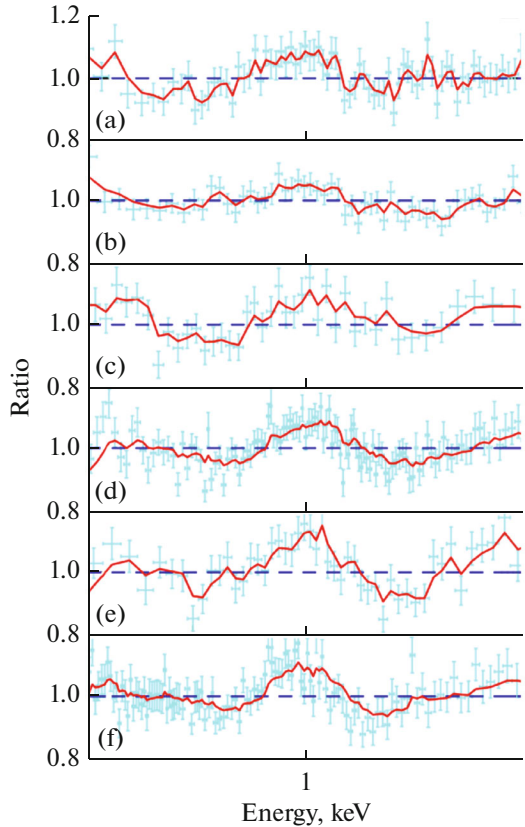

**Fig. 6.** Residuals, remaining after dividing the observed X-ray spectra of ULXs by the TBABS*(DISKBB+NTH-COMP) model (multicolor disk+comptonization). The objects NGC 1313 X-1, Ho IX X-1 (hard), Ho II X-1 (soft), NGC 55 ULX-1 (super soft), NGC 6946 X-1, NGC 5408 X-1 (soft) are shown from top to bottom. All residuals have a similar shape with a hump near 1 keV despite the difference in the object types. Residuals of this kind indicate the presence of features in the spectrum that are not taken into account by simple continuum models (Fig. 1 from the paper Middleton et al. (2015b)).

By modeling the multicolor funnel, it was predicted that if the internal parts of the SS 433 funnel were observed directly (face-on), a powerful hump at an energy of about 1 keV would appear in the spectrum, along with small, very wide $(0.1c–0.3c)$ blue-shifted absorptions. As soon as the ULX spectra with high signal-to-noise ratio were obtained, such humps were actually detected (Middleton et al. 2015b). Fig. 6 shows the residuals of the spectra of the six most studied ultraluminous X-ray sources. It can be seen that all of them have approximately the same shape, indicating that these residuals are not random in nature, but are the result of absorption in the funnel, which is not taken into account by simple continuum models. However, the shape of the residuals may vary slightly depending on the type of spectrum (SUL or HUL) and the inclination of the supercritical disk (Pinto et al., 2017a, 2020).

The observed X-ray luminosity of SS 433 is $L_X \sim 10^{36}$ erg s$^{-1}$. Most of this radiation comes from jets which have a temperature of jet's visible base of 17 keV (Brinkmann et al., 2005). However, approximation of the observed spectrum with only jet model yields residuals that systematically increase toward high energies indicating the need for some additional spectral component (Brinkmann et al., 2005; Khabibullin et al., 2016; Medvedev and Fabrika, 2010). In the work Medvedev and Fabrika (2010) we analyzed the spectra of SS 433 obtained with the XMM-Newton telescope, applying the model of an adiabatically expanding radiatively cooling jet, we found that an additional component may be radiation from the inner parts of the disk reflected towards the observer by the funnel walls. The additional component becomes statistically significant in the energy range of more than 3 keV. Reflection occurs on a relatively cold $(T \sim 0.1$ keV) but highly ionized plasma (ionization parameter $\xi \sim 300$). The reflected hard radiation has power-law spectrum with the exponent $\Gamma \sim 1.5–2$. A significant contribution of reflection to the observed SS 433 spectrum is also indicated by the prominent 6.4 keV fluorescence line.

Using the data collected by the RXTE/ASM X-ray telescope from 1996 to 2005, SS 433 spectroscopy was performed, and Filippova et al. (2006) found that the difference between the energy spectra obtained at different precession and orbital phases demonstrates the effect of strong photoabsorption near the optical star due to its wind. Due to the presence of wind, the size of the star obtained from the analysis of X-ray eclipses can be significantly larger than the Roche lobe. Taking this effect into account, the jet temperature and the distance from the visible part of the jet to the compact object were more accurately measured: $2–3 \times 10^{11}$ cm.

Figure 7 shows high-resolution spectra obtained for SS 433 using the HETGS (High Energy Transmission Grating Spectrometer) instrument on board the Chandra observatory (Marshall et al., 2002). The spectra show many emission lines of highly ionized elements (Fe XXV, S XVI, Si XIV, Si XIII) with a Doppler shift to the blue and red sides. These lines originates from the jets. The emission lines are broadened to $FWHM \approx 1700$ km s$^{-1}$, but their width does not depend on the radiation temperature. This means that the jet gas moves along a strictly ballistic trajectory with an opening angle of $\theta = 1°.23 \pm 0°.06$. The Doppler shifts of the blue lines are also independent of temperature. From this, we can conclude that the hottest parts of the jet, observed first after the jet appears above the funnel's photosphere, have already been accelerated to the maximum speed, which is $0.2699 \pm 0.0007c$ of the speed of light. Based on the self-consistent jet model (Marshall et al., 2002), the temperature—from $5 \times 10^6$ K and up to $1 \times 10^8$ K in different parts of the jet—and the kinetic luminosity $L_j \sim 3.2 \times 10^{38}$ ergs$^{-1}$ were estimated. The spectral continuum in the region of less than 3 keV is almost entirely produced by bremsstrahlung radiation of jet plasma.

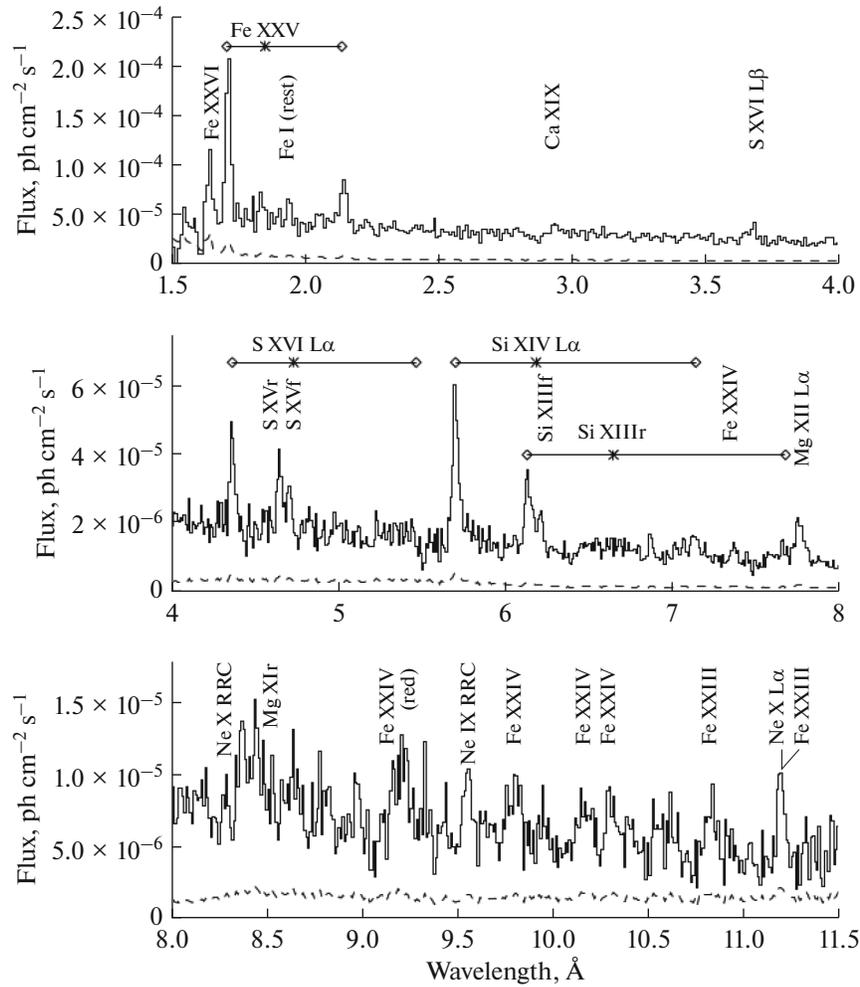

**Fig. 7.** SS 433 spectra from the article Marshall et al. (2002) obtained using the high-resolution X-ray spectrograph HETGS of the Chandra. The sources of almost all radiation in the presented range are relativistic jets (blue and red). Horizontal lines with rhombuses at the ends mark the Doppler shifts of the lines of the blue and red jets, asterisks are the position of the lines in the laboratory frame. The dashed line corresponds to statistical uncertainty. The emission lines are resolved, and their widths indicate that the plasma is collisional.

Several years ago, relativistic lines were also detected in the high-resolution spectra of ultraluminous X-ray sources. In Pinto et al. (2016), where the spectra of NGC 5408 X-1, NGC 1313 X- 1 and NGC 6946 X-1 objects obtained with the RGS (Reflection Grating Spectrometer) of the XMM-Newton observatory were studied, it was shown that the first two objects, in addition to the resting emission lines, also have absorption lines blue-shifted by $0.2c$–$0.25c$ (Fig. 8). The presence of such lines indicates that the line of sight is crossed by gas flows moving towards the observer at relativistic speeds; they are called *ultrafast outflows (UFO)*.

Later, ultrafast outflows were also found in the ultraluminous pulsar NGC 300 ULX-1 (Kosec et al., 2018) and in the ultraluminous supersoft source NGC 55 ULX (Pinto et al., 2017a). It is noteworthy that in the latter case, both the absorption and emis-sion lines were blue-shifted by $0.01c$–$0.20c$. This fact confirms the relationship between the spectral type of the object and the inclination of the disk to the line of sight: the objects NGC 5408 X-1 and NGC 1313 X-1, observed along the funnel axis and having the types SUL and HUL, respectively (Sutton et al., 2013), have UFO only in absorptions, but the object NGC 55 ULX, observed from the side—both in absorptions and in emissions (Fig. 4).

In Pinto et al. (2020), the wind photoionization balance was calculated for nine objects, including ULXs and ULSs. It has been shown that the wind is usually in a thermally stable equilibrium, but its stability and spectrum are affected by changes in the accretion rate and disk inclination. In particular, this can explain the variation in the shape of residuals in many objects in the region of 1 keV (Fig. 6) over time. A cor-relation was also found between the object's hardness, wind velocity, and the wind ionization parameter ξ.

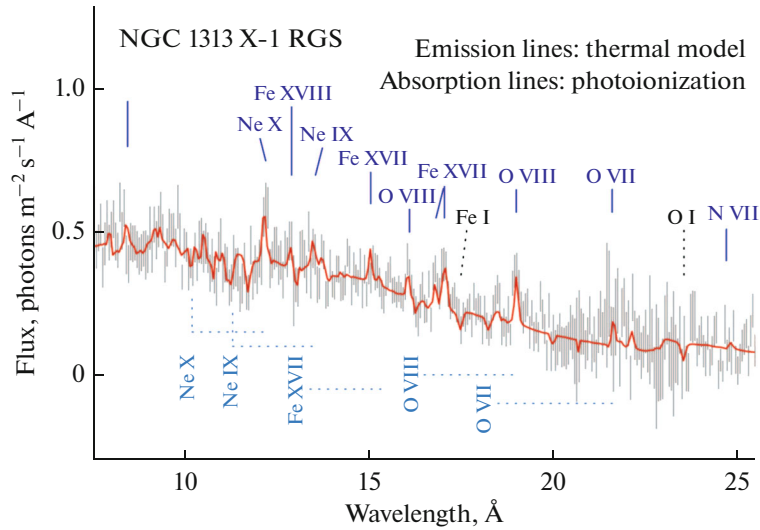

**Fig. 8.** High-resolution spectra of NGC 1313 X-1 from the article Pinto et al. (2017a) obtained using the XMM-Newton telescope's RGS spectrograph. The main lines are signed, and the dotted line shows the offset of the lines in relation to the rest frame.

### 2.2. Ultraluminous X-Ray Pulsars

Perhaps one of the most unexpected discoveries was the discovery in 2014 by Matteo Bachetti's group of coherent pulsations of the ultraluminous X-ray source M 82 X-2 (Bachetti et al., 2014), which was already a well-known object at that time. Its pulsations are modulated by orbital motion with a period of about 2.5 days and have an average period of 1.37 s (Fig. 9). As in the case of ordinary X-ray pulsars, in M 82 X-2 pulsations occur due to accretion of matter onto a rotating magnetized neutron star. The peak luminosity of the source is $L_X = 1.8 \times 10^{40}$ erg s$^{-1}$, which exceeds the Eddington limit by more than 100 times (for the mass of a neutron star, 1.4 $M_\odot$).

By 2020, quite a lot of ultraluminous pulsars have become known. In addition to M 82 X-2, these are NGC 7793 P13 (Fürst et al., 2016; Israel et al., 2017b), NGC 5907 ULX-1 (Fürst et al., 2017; Israel et al., 2017a), NGC 300 ULX-1 (Carpano et al., 2018), M 51 ULX-7 (Rodriguez Castillo et al., 2020), M 81 X-6 (Jithesh et al., 2020) and NGC 1313 X-2 (Sathyaprakash et al., 2019). The brightest is the pulsar NGC5907 ULX-1, its luminosity reaches the order of $10^{41}$ erg s$^{-1}$ (see Table 1), which puts it in the category of hyperluminous sources. Most of the objects have periods in the range from a fraction to tens of seconds.

**Table 1.** Parameters of ultraluminous X-ray pulsars

| Object | $L_{\mathrm{X,max}}$, erg s$^{-1}$ | $P$, s | $\dot{P}_{\max}$, s$^{-1}$ | References |
|---|---|---|---|---|
| M 82 X-2 | $1.8 \times 10^{40}$ | ~1.37 | $-2.7 \times 10^{-10}$ | Bachetti et al. (2014) |
| NGC 7793 P13 | ~$10^{40}$ | ~0.43 | $-3 \times 10^{-11}$ | Fürst et al. (2016); Israel et al. (2017b) |
| NGC 5907 ULX-1 | ~$10^{41}$ | ~1.1 | $-8 \times 10^{-10}$ | Fürst et al. (2017); Israel et al. (2017a) |
| NGC 300 ULX-1 | $4.7 \times 10^{39}$ | ~31.5 | $-5.6 \times 10^{-7}$ | Carpano et al. (2018) |
| M 51 ULX-7 | $7 \times 10^{39}$ | ~2.8 | $-10^{-9}$ | Rodriguez Castillo et al. (2020) |
| M 81 X-6 | $3.6 \times 10^{39}$ | 2681 | — | Jithesh et al. (2020) |
| NGC 1313 X-2 | $2 \times 10^{40}$ | ~1.5 | $-3.3 \times 10^{-8}$ | Sathyaprakash et al. (2019) |
| M 51 ULX-8 | $4.8 \times 10^{39}$ | — | — | Brightman et al. (2018); Middleton et al. (2019) |
| CXOU J073709.1+653544 | ~$10^{39}$ | ~18 | $-1.1 \times 10^{-7}$ | Trudolyubov (2008) |
| SMCX-3 | $2.5 \times 10^{39}$ | ~7.8 | $-7.4 \times 10^{-10}$ | Townsend et al. (2017); Tsygankov et al. (2017) |
| RXJ0209.6-7427 | ~$2 \times 10^{39}$ | ~9.3 | $-1.75 \times 10^{-8}$ | Chandra et al. (2020); Vasilopoulos et al. (2020b) |
| SwiftJ0243.6+61241 | ~$2 \times 10^{39}$ | ~9.8 | ~$2.2 \times 10^{-8}$ | Doroshenko et al. (2018, 2020); Tao et al. (2019), Zhang et al. (2019) |

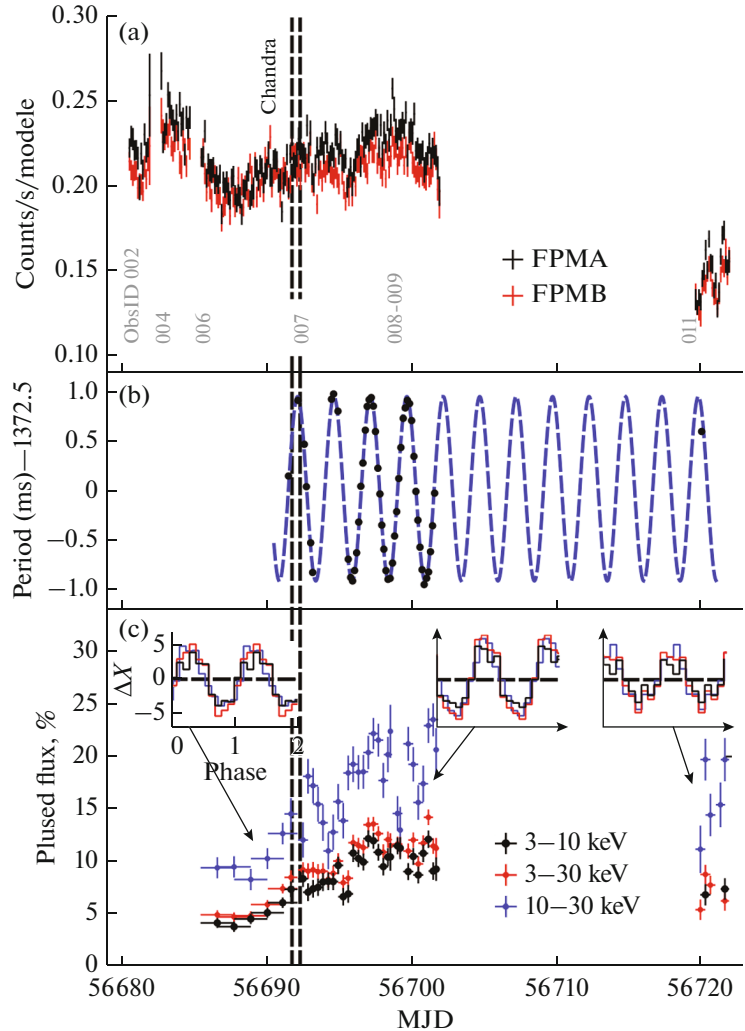

**Fig. 9.** Figure from the work Bachetti et al. (2014): discovery of pulsations in the object M 82 X-2. Panel (a): the light curve of the object obtained using the NuStar telescope; data from two identical detectors are shown in black and red. Panel (b): the pulsation period (black dots) and its modulation by orbital motion, reconstructed using the best sinusoidal ephemerides (blue dashed line). The pulsation period is 1.37 s, the orbital period is 2.5 days. Panel (c): the fraction of pulsed radiation to total for three bands. The inserts show the pulse profiles. Dashed vertical lines indicate the moments of observations with the Chandra telescope.

M 81 X-6, which has a period of 2681 s, differs sharply from them, but the statistical significance of this period is not very high (just above 95%) (Jithesh et al., 2020). The fraction of pulsed emission (pulsed fraction) for all ULXPs increases with increasing energy; for the object NGC 7793 P13 at of 10 keV, it reaches 50% (Israel et al., 2017b).

Another confirmed ULX with a neutron star is M 51 ULX-8 (Brightman et al., 2018). This source does not show pulsations, but a cyclotron line has been found in the X-ray spectrum—this clearly indicates the presence of a strong magnetic field in this object. However, Brightman et al. (2018) notes that the shape of the found line differs from that observed in ordinary pulsars: the line of M 51 ULX-8 is narrower and has no harmonics. It is assumed that either this line can be associated with proton transitions (Brightman et al.,

2018), and not with electronic ones, or the magnetic field can have a significant multipole component (Middleton et al., 2019).

Additionally, several transient pulsars, that have ever shown peak luminosities above $10^{39}$ erg s$^{-1}$ are also considered as ULXPs: CXOU J073709.1+653544 in the galaxy NGC 2403 (Trudolyubov, 2008), as well as the pulsars SMCX-3 (Townsend et al., 2017; Tsygankov et al., 2017) and RX J0209.6-7427 (Chandra et al., 2020; Vasilopoulos et al., 2020b) in the Magellanic Clouds and Swift J0243.6+61241 (Doroshenko et al., 2018; Tao et al., 2019) in our Galaxy. Although such luminosities only barely fall under the formal definition of ULX, nevertheless, given that accretion goes to neutron stars, they exceed the Eddington limit by a factor of ten or more. Unlike objects of the first group—standard ULXPs, which can accrete in the super-Eddington regime for a long time—transient pulsars reach super-Eddington luminosities only at relatively short outburst times. During an outburst, there is a sharp rise and then an exponential decline in luminosity, lasting several months (see light curves, for example, in Tao et al. (2019); Townsend et al. (2017)).

It is assumed that transient ULXPs are binary systems with Be donors in which the neutron star moves in an elliptical orbit (King and Lasota, 2019). Be stars are a special type of stars that form a circumstellar disk due to rapid rotation. A neutron star draws matter from this disk during the periastron passage; but once every few revolutions, the flow of matter can increase dramatically, for example, due to the fact that the plane of the disk and the plane of the orbit do not coincide (Martin et al., 2014). The assumption about a sharp increase in the accretion rate during outbursts is confirmed by the fact that at these moments the spin-up rate of the neutron star increases abruptly (the derivative $\dot{P}$ increases modulo) (Chandra et al., 2020; Zhang et al., 2019). In general, the high spin-up rates—several orders of magnitude higher than that of ordinary pulsars—is inherent to some extent in all ULXPs (see Table 1).

The detection of pulsations clearly showed that some of the ultraluminous X-ray sources are accreting neutron stars. This raises the question: how widely are neutron stars represented in the ULX population? Analysis of a sample of 15 most studied objects with a large number of observations showed that there are no pulsations in them (Doroshenko et al., 2015). However, in confirmed ULXPs pulsations manifest themselves not in every observation. Thus, in M 82 X-2, pulsations were found only in observations of 2014; in earlier data (the source has already been observed several dozen times before), there are no pulsations (Bachetti et al., 2014). This was confirmed by a subsequent more careful analysis of old observations. After the discovery of pulsations, extensive monitoring of this object began: in 2015 and 2016 the NuStar observatory made 15 new observations. Of these, weak pulsations were detected only in one observation (Bachetti et al., 2020). The situation is similar with other ULXPs. This is due to the small (and varying from observation to observation) fraction of the pulsating part of the radiation (Fig. 9c), as well as to the modulation of pulsations by orbital motion, which smears out the peak corresponding to the pulsation frequency in the power spectrum. The article Rodriguez Castillo et al. (2020) provides interesting statistics, according to which out of about 300 known ULXs (Earnshaw et al., 2019), 15 objects have sufficient accumulation to detect pulsations, and in 25% of them pulsations have been already found.

As for other properties, besides the presence of pulsations, ultraluminous pulsars do not differ much from other ULXs. In general, they have more hard spectra, but their shape is quite similar to the type BD or HUL (Pintore et al., 2017; Walton et al., 2018). Also, transient and standard ULXPs show a sharp drop in luminosity of 100 or more times, which is not typical for most ULXs (see below). But these differences are not enough to identify a neutron star by indirect signs.

In this regard, we can conclude that the question of the ratio of black holes and neutron stars among ULXs is still far from an unambiguous solution. It has been suggested that almost all of the known ULXs of known objects may be neutron stars (King and Lasota, 2016; King et al., 2017; Middleton and King, 2017; Mushtukov et al., 2015; Walton et al., 2018) (however in many of them pulsations can be suppressed, see below), but there are more cautious opinions. Evolutionary calculations show that, in general, the binary systems with neutron stars should be 50−100 times more numerous than systems with black holes (Belczynski and Ziolkowski, 2009). However, if the lifetime of such systems in the ULX mode is very short, or it requires some extreme conditions (for example, the presence of very strong magnetic fields), then the number of ULXs with neutron stars may be relatively small.

Being accreted onto a magnetized neutron star, the matter of the accretion disk, starting from the radius $r_m$, will be captured by the magnetic field and then forced to move along its lines of force. This radius, which determines the size of the pulsar's magnetosphere, depends on the magnitude of the magnetic field, the accretion rate, and the mass of the neutron star; it is called the *Alfven radius*. The magnetic field lines forward the flow of matter to the poles of the neutron star, where the socalled *accretion column* is formed near each pole (Basko and Sunyaev, 1975, 1976). Due to the fact that the matter is held by the magnetic field, as well as due to the non-spherical geometry of the flow, the luminosity of the accretion column can exceed the Eddington level by several times (Basko and Sunyaev, 1976). Modern models (Chashkina et al., 2017, 2019; Mushtukov et al., 2015, 2017, 2018b, 2019a) show that magnetic fields of $B \gtrsim 10^{14}$ G are required to provide luminosity of the order of $10^{40}$ erg s$^{-1}$. These are very strong fields, they are two orders of magnitude higher than the fields observed in "ordinary" X-ray pulsars. Nevertheless, several neutron stars with such fields are known in our Galaxy; such objects are called *magnetars*. In Mushtukov et al. (2015), it is noted that the luminosity of the order of $10^{40}$ erg s$^{-1}$ may be the physical limit for pulsars, which explains the presence of a break in the luminosity function of X-ray sources at this value (Mineo et al., 2012).

In the works by Chashkina et al. (2017); Mushtukov et al. (2015, 2017) it is assumed that $r_m > r_{sp}$, i.e. there is no supercritical region in the disk, and the matter is being captured immediately from the standard disk. However, due to the high rate of accretion, the flow of matter moving along the field lines is optically thick (Mushtukov et al., 2017, 2019a). The shell formed by this flow (so-called accretion curtain) must have the spectrum of a multicolor black body with a temperature of several kiloelectronvolts. Thus, it is potentially possible to describe the observed X-ray spectra of ULXs. These models also explain the absence of cyclotron lines in the spectra of ultraluminous pulsars and the differences in the pulsation profiles of ULXPs and ordinary pulsars (Mushtukov et al., 2017, 2019a).

The most serious argument in favor of the presence of strong fields in ULXPs is the discovery of the bimodal luminosity distribution of the ultraluminous pulsar M 82 X-2 (i.e., the source is often either bright or weak, but rarely in intermediate states), which can be interpreted as the result of the so-called *propeller effect* (Tsygankov et al., 2016b). The possibility of the presence of this effect, predicted in the 70s by Illarionov and Sunyaev (1975) (see also Lipunov, 1987; Ustyugova et al., 2006), has recently been reliably confirmed by observations of two pulsars of our Galaxy (Tsygankov et al., 2016a). The propeller effect is that if at the radius of the magnetosphere $r_m$ the angular velocity of the matter forming the accretion disk is lower than the angular velocity of the neutron star spin, then accretion stops, because centrifugal forces begin to prevent the fall of the matter. As the accretion rate decreases, the radius of the magnetosphere increases as $r_m \propto \dot{M}^{-2/7}$ and the Kepler angular velocity at this radius $\Omega_K(r_m)$ become smaller, so under certain intermediate conditions, even a small decrease in $\dot{M}$ can stop accretion and the source will completely go out. It was shown that in order for the propeller effect to work at the observed luminosities of M82 X-2, fields of the order of $10^{14}$ G are required (Tsygankov et al., 2016b). An extensive search for the propeller effect among ULXs was undertaken (Earnshaw et al., 2018; Song et al., 2020). Out of several hundreds of analyzed objects, 25 were found to have variability with an amplitude of more than 10 times, and signs of a bimodal distribution were found in 17 objects.

Another important argument in favor of the presence of magnetar-like magnetic fields in M 82 X-2 is the detection of a spindown in the observations of 2014–2016 (Bachetti et al., 2020). The alternation of spin-up and spin-down moments in the pulsar life may indicate that its spin period is close to equilibrium. The spin up occurs due to the fact that the accreted matter brings angular momentum. At those times when accretion stops, the pulsar's rotation begins to slow down due to the losses by radiation that the neutron star emits being a rotating magnetic dipole. The magnetic field estimated from these considerations is approximately $3 \times 10^{14}$ G (Bachetti et al., 2020), which corresponds to the value obtained from the relations that take into account the propeller effect.

In addition to models with strong magnetic fields, another class of models have been proposed the presence of magnetar fields is not required to explain the ULXP phenomenon, and large observed luminosities is attributed to high collimation (King and Lasota, 2016, 2019, 2020; King et al., 2017; Kluzniak and Lasota, 2015; Middleton and King, 2017; Walton et al., 2018), accretion flow geometry (Kawashima and Ohsuga, 2020; Kawashima et al., 2016) or other mechanisms (Takahashi et al., 2018). The last work shows that supercritical accretion is possible even on a neutron star that has no magnetic field at all. Based on the MHD calculations, Takahashi et al. (2018) showed that in this case an accretion disk should have a structure generally similar to that described in the Introduction of this review (i.e., a conical funnel with $H \sim R$ filled with rarefied gas moving at relativistic velocities), but having a more powerful wind. Near the surface of a neutron star, the gas stops and a "cushion" consisting of matter and radiation is formed. The excess of the Eddington luminosity is achieved as a result of the fact that the radiation pressure is balanced by the sum of the centrifugal and gravitational forces (Takahashi et al., 2018).

In the papers Kawashima et al. (2016) and Kawashima and Ohsuga (2020), MHD calculations of the accretion column are presented for the case of weak, $B \sim 10^{10}$ G, and average, $B \sim 10^{12}$ G (observed in most ordinary pulsars), fields, respectively. In the first case, the gas can still move slightly across the magnetic field lines, and the column appears to be filled with matter. In the second case such movements are already completely "forbidden," the column looks like a multi-layered (onion-like) hollow cone. The space between the layers is filled with very rarefied gas, which moves in the opposite direction respect to the main flow (ejected from the system). Since that radiation can freely escape the accretion column through its side walls, the Eddington limit can be exceeded by several orders of magnitude in the first model and up to 30 times in the second one. Such a huge efficiency in the case of weak fields is due to the fact that in a filled column near the surface of a neutron star, powerful shock waves arise, effectively converting the kinetic energy of the incident matter into radiation.

The size of the magnetosphere decreases as the magnetic field weakens, so in models with weak and medium fields (King and Lasota, 2019, 2020; King et al., 2017; Walton et al., 2018) gas capturing by the magnetic field occurs in the supercritical zone of the disk, i.e. $r_m \lesssim r_{sp}$. The observed fraction of the pulsating part of the radiation in this case should depend on the ratio between these two radii. Since the pulsations will be smeared out due to the scattering of radiation in the funnel of the supercritical disk and since the funnel itself is also a source of non-pulsating radiation, it is assumed that $r_{sp}$ must exceed the radius of the magnetosphere by no more than 2–3 times in order to reli-ably observe the pulsations (King and Lasota, 2019; King et al., 2017; Walton et al., 2018). At $r_m \ll r_{sp}$, the fraction of pulsating radiation will be too small to be detected, so it is suggested by (King and Lasota, 2016, 2020; King et al., 2017; Middleton and King, 2017) that many ULXs whose pulsations are not yet found may also include neutron stars.

Also, the funnel is likely to collimate the radiation of the central source, and the degree of collimation may depend on the dimensionless accretion rate (King, 2009) (which, at the same accretion rate in absolute units, is greater for neutron stars than for black holes). This partly explains the observed large luminosities of neutron stars, but too high collimation factors look doubtful. In addition, the sinusoidal pulsation profiles observed in ULXPs, which are smoother compared to the profiles of ordinary pulsars, can be associated with partial scattering of radiation in the funnel (Fig. 9). In this case, the specific shape of the pulsation profile and the pulsed fraction should depend on the mutual orientation of the rotational axis of the pulsar, its magnetic axis and the axis of the accretion disk funnel in respect to each other and to the observer's line of sight (King and Lasota, 2020). It is noted that if the axis of rotation is oriented approximately along the line of sight, the pulsations will be suppressed even at $r_m \sim r_{sp}$. This situation may occur in the source M 51 ULX-8, which has a cyclotron line, but no pulsations (King and Lasota, 2020).

It is important to emphasize that both classes of models have their own weaknesses. Strong magnetic fields of the magnetar level are subject to rapid decay, especially at such high rates of accretion. Although in our Galaxy, a candidate for magnetars with an age of 2.4−5.0 Myr was found (the object 4U 0114+65), and a possible scenario for its formation has been proposed (Igoshev and Popov, 2018), nevertheless this class of ULXP models requires the simultaneous coincidence of two extreme conditions: the presence of strong fields and permanently high accretion rates, which means that if such objects do exist, they most likely cannot be numerous. The results of population synthesis show that in binary systems with neutron stars, the rate of matter transfer necessary to provide the luminosity of the system at the ULX level occurs only at ages from 100 Myr and more (Wiktorowicz et al., 2017). Also against the presence of strong magnetic fields in ULXPs, King and Lasota (2019) argues that all known magnetars are isolated objects, while in binary systems the fields of neutron stars do not exceed $10^{13}$ G. To this, however, it may be objected that cyclotron lines, by which the fields can be measured directly, are found only in a relatively small number of pulsars. And the absence of cyclotron lines in the spectra of other objects may be due, among other things, to the fact that the fields of these objects are too strong, and the lines are beyond the energy range available for measurements.

In models with collimation, the extreme parameter is only one—the accretion rate. They explain why most ULXs do not show pulsations (if they are neutron stars, then due to the weak field, their magnetospheres may be too small compared to the size of the supercritical disk funnel), and also why the other properties of ULXs, among which there may still be black holes, and ULXPs turned out to be so similar (the properties are determined more by the supercritical disk than by the magnetosphere). However, models with collimation fails to explain the nature of transient ULXPs. During the outburst, the luminosity of these objects varies over a wide range, but there is no indication that the object has moved from a low collimation state to a high collimation state. There are also problems with the description of those objects that have a significant pulsed fractions at a colossal luminosity—in particular, this concerns the pulsar NGC 5907 ULX-1, which has a pulsed fraction of 12−20% at a luminosity of more than $10^{41}$ erg s$^{-1}$. The pulsating component of the emission can hardly be strongly collimated, since with multiple scattering in a narrow and long funnel, the pulsations must be washed out. However, even for models with magnetar fields, this object is difficult to describe. In (Israel et al., 2017a), it is reported that fields of the order of $5 \times 10^{15}$ G are required to explain the luminosity of NGC 5907 ULX-1, but with such fields, the source must be in the propeller mode constantly, and accretion cannot occur. To overcome this difficulty, an explanation was proposed that the field can be multipole, and the radiation might be collimated by 7−25 times (Israel et al., 2017a).

## 2.3. Variability in the X-Ray Range

The variability of the radiation of astronomical objects can be divided into periodic and stochastic. The most obvious example of periodic variability is the variability associated with orbital motion. SS 433 is an eclipsed binary system. So once per the orbital period $P_{orb} = 13.082$ days, when the donor star eclipses the accretion disk, a sharp decline in X-ray flux occurs, and twice per orbital period—a noticeable decrease of the optical flux (i.e. both when the disk is eclipsed by the star and the star by the disk, the contribution of the star is approximately 10% of the total optical radiation of the system, see below). In addition to the orbital period, SS 433 also has a precession $P_{pr} \approx [162.2; 162.5]$ days and nutation $P_{nut} = 6.28$ days periods (nodding of the jets due to tidal forces) (Fabrika, 2004).

As for ULXs, we found in the literature mention of five objects for which successful detection of eclipses was reported: these are the source in the Circinus galaxy (Circinus Galaxy X1, CG X-1 (Qiu et al., 2019)) and four sources in the M 51 galaxy: CXOM51 J132940.0+471237, CXOM51 J132939.5+471244 (Urquhart and Soria, 2016b), CXOM51

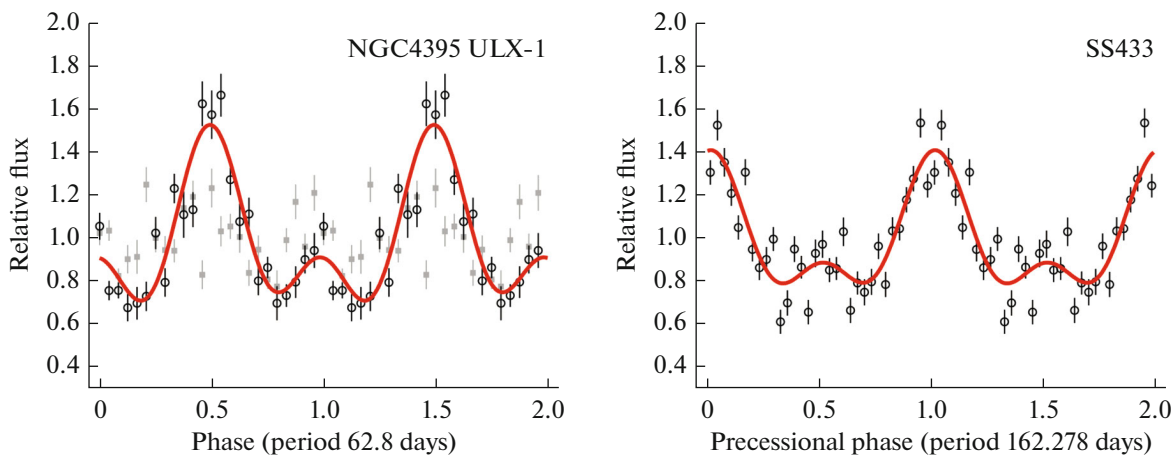

**Fig. 10.** (a) The X-ray epoch folded light curve of the source NGC 4395 ULX-1 from the work Vinokurov et al. (2018), corresponding to the period $62.8 \pm 0.3$ day. Black indicates the points of the object, gray—points of the background. The fluxes are given in relative units (normalized to the average level), but in reality the background signal is more than 50 times weaker than the object signal. The red line shows the approximation of the observed points to the two harmonics of the Fourier series. (b) For comparison, the precession phase curve of SS 433, constructed from RXTE/ASM data, is shown.

J132943.3+471135 and CXOM51 J132946.1+471042 (Wang et al., 2018). For CGX-1, the orbital period of the system $P \approx 7.2$ hours was found and its derivative was measured $\dot{P}/P \sim 10^{-6} \, \mathrm{yr}^{-1}$ (the period increases) (Qiu et al., 2019). In CXOM51 J132946.1+471042, an orbital period of $52.75 \pm 0.63$ hours was determined from the Chandra series of observations and the eclipse depth was measured—of the order of 22% (Wang et al., 2018).

Some other objects also showed sharp dips in the X-ray flux, similar to eclipses. In particular, a "period" of 115 days was found for NGC 5408 X-1 (Strohmaye, 2009). However, it was later found that these dips sometimes disappear and, accordingly, cannot be real eclipses (Grisé et al., 2013; Pasham and Strohmayer, 2013a). Most likely, they are associated with wind clouds crossing the line of sight (see below).

A number of objects were found to have superorbital periods associated, most likely, with the precession of the accretion disk: NGC 4395 ULX-1 of the order of 63 days (Vinokurov et al., 2018), Holmberg XI X-1 about 266 days, NGC 1313 X-1 about 212 days, in ultraluminous pulsars NGC 5907 ULX-1 of the order of 78 days (Walton et al. 2016), M 82 X-2—60 days (Brightman et al., 2019), M 51 ULX-7—39 days (Vasilopoulos et al., 2020a), NGC 1313 X-2—158 days (Weng et al., 2018), as well as some other objects (Lin et al., 2015; Weng et al., 2018). Fig. 10 shows the phase curve of NGC 4395 ULX-1, plotted from 226 points of Swift/XRT observations conducted from 2005 to 2015. The figure shows that, in addition to the main maximum, the phase curve also has a secondary maximum, which makes it very similar to the precession curve of SS 433 (Atapin and Fabrika, 2016).

Stochastic variability is more or less common to all accreting systems (Frank et al., 2002), and ultraluminous X-ray sources are no exception. Almost all ULXs on the time scale months—years change their brightness at least by factors 3—5 (Pintore et al., 2014; Sutton et al., 2013). Fig. 11 shows the light curve of IC 342 X-1. Some objects show more significant variability—by a factor of 100 or more. These are mostly ultraluminous pulsars (for example, M 82 X-2 (Tsygankov et al., 2016b), see also references in the previous section), but not all of these objects show pulsations, for example, sources in the galaxies M 83 (Soria et al., 2012), M 86 (van Haaften et al., 2019), NGC 925 (ULX-3) (Earnshaw et al., 2020), UGC6456 (Brorby et al., 2015; Vinokurov et al., 2020) and others (Earnshaw et al., 2018).

At shorter times, within a single continuous observation (the longest ones are usually less than $10^5$ s), a relationship was found between the variability and the type of spectrum of the object. To quantify the variability, the characteristic *fractional rms variability*, $F_{\mathrm{rms}}$ is often used—the standard deviation of the light curve samples normalized to the average flux, minus the contribution of the (Poisson) measurement noise (Vaughan et al., 2003). It turned out that the fractional mass variability is higher for sources with a soft spectrum and reaches 40% (Pintore et al., 2014; Sutton et al., 2013). However, if we compare the contributions to the variability of the soft and hard parts of the spectrum of soft sources, it turns out that the largest contribution is made by the hard (De Marco et al., 2013; Hernández-García et al., 2015; Pintore et al., 2014; Sutton et al., 2013).

Ultraluminous supersoft sources are even more variable. Even on time scales of several hours, their

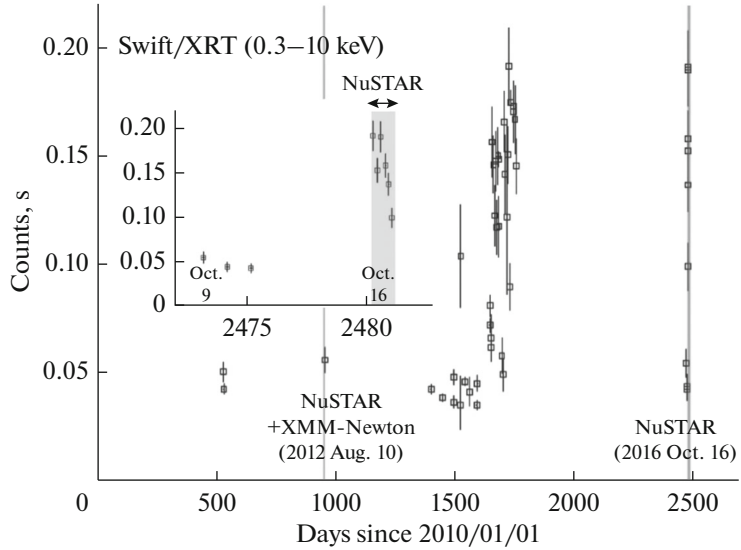

**Fig. 11.** The light curve of the object IC 342 X-1 from the paper Shidatsu et al. (2017), obtained using the Swift/XRT X-ray telescope. Gray stripes, as well as large in the inset, show the times of simultaneous observations of Swift+XMM-Newton and Swift+NuStar. The figure shows that the source changed its luminosity by about 3 times in less than a day (see also Fig. 3).

X-ray light curves show sharp dips lasting tens of minutes (Feng et al., 2016; Pinto et al., 2017a; Urquhart and Soria, 2016a). In addition to dips, there are also flares (for example, the light curves of the source M 101 ULS (another name M 101 ULX-1) in the work Urquhart and Soria (2016a)), and here the variability in the hard range is higher also.

This relationship between $F_{rms}$ and the spectrum type is in good agreement with the scheme shown in Fig. 4. The softer the spectrum, the greater inclination of the fuel, and the more often the line of sight is crossed by optically thick clumps of wind escaping the funnel wall (Middleton et al., 2015a; Pinto et al., 2017a; Sutton et al., 2013; Urquhart and Soria, 2016a). These clumps obscure the inner hot parts of the funnel that emit the hard spectrum, resulting in dips in the light curve. In ULSs observed almost along the funnel wall, the hot parts of the funnel can be covered by gas clouds almost constantly, and when a "window" is formed in them, a flare is observed in the hard range. Efficiency of this mechanism of variability caused by opaque clumps of the supercritical disk wind has been repeatedly confirmed by MHD modeling (Takeuchi et al., 2013, 2014).

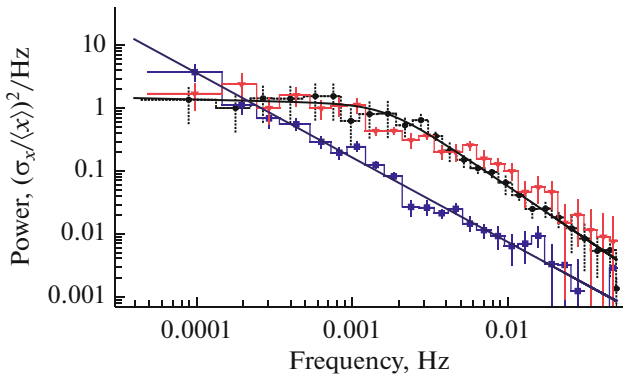

**Fig. 12.** Power spectra of SS 433 in the range 2−20 keV according to RXTE/PCA data for different orientations of the supercritical disk (precession phases). Black shows the power spectrum corresponding to such a precession phase, when the funnel is maximally opened towards the observer, blue—when the disk is observed from the edge (the funnel is completely closed), red—an intermediate orientation. The "black" and "red" power spectra have an explicit flat section $P \propto f^0$ at frequencies below $10^{-3}$ Hz.

Another mechanism of variability is variations in the accretion rate due to viscosity fluctuations at different disk radii. This idea was proposed by Lyubarskii (1997) to explain power-law power spectra and refined in articles by other authors (Arévalo and Uttley, 2006; Ingram and Done, 2011; Ingram and van der Klis, 2013; Kotov et al., 2001; Mönkkönen et al., 2019; Mushtukov et al., 2018a, 2019b; Revnivtsev et al., 2009; Titarchuk et al., 2007). Power-law power spectra are observed in many X-ray binaries of our Galaxy, although their detailed shape (exponents, break frequencies, etc.) depends on the specific accretion state (Belloni, 2018; McClintock and Remillard, 2006).

SS 433 also demonstrates a power-law power spectrum with an exponent of about 1.5 (Revnivtsev et al., 2004, 2006). However, a more detailed analysis (Atapin et al., 2015) showed that the shape of its power spectrum depends on the precession phase (Fig. 12). The power spectrum has a purely power-law form only when the funnel of the supercritical disk is closed to the observer. Further, as the precession phase changes and the observer looks deeper into the funnel, a flat section or flat-topped noise (FTN) appears in the power spectrum. We believe that this form of the power spectrum (with a flat section) is a manifestation of supercritical accretion in this object.

ULX power spectra were studied in Agrawal and Nandi (2015); Caballero-García et al. (2013a, b); De Marco et al. (2013); Heil et al. (2009); Hernández-García et al. (2015); Pasham and Strohmayer (2012, 2013b); Pasham et al. (2015); Rao et al. (2010); Strohmayer and Mushotzky (2003); Strohmayer et al. (2007). The objects M 82 X-1, NGC 5408 X-1 and NGC 6946 X-1 are the most studied (and most variable at frequencies greater than $10^{-3}$ Hz). M 82 X-1 is the first ULX to have quasiperiodic oscillations (QPO). Quasiperiodic oscillations are observed in many X-ray transients of our Galaxy; their frequencies change as the outburst progresses, but almost never fall below 0.1 Hz (Belloni, 2018; Motta et al., 2011). For NGC 5408 X-1 and NGC 6946 X-1, QPO peaks are observed at frequencies $0.01-0.04$ Hz (Caballero-García et al., 2013a; De Marco et al., 2013; Pasham and Strohmayer, 2012; Rao et al., 2010).

In Atapin et al. (2019), we investigated the variability of five ULXs, in which QPOs were found: NGC 5408 X-1, NGC 6946 X-1, M 82 X-1, NGC 1313 X-1 and IC 342 X-1. Their power spectra are shown in Fig. 13. It can be seen that all of them have a rise towards low frequencies and a flat section similar to that observed in SS 433 (Fig. 12), and the level of the flat section is higher the lower the QPO frequency.

We also studied the relationships between QPO frequencies, fraction mass variability, X-ray luminosity, and the hardness of the spectrum of the objects (Atapin et al., 2019). All these parameters for each of the studied sources vary from observation to observation. It was found that $F_{rms}$ is anticorrelated with the QPO frequency (Fig. 14a). In this case, the objects NGC 5408 X-1, NGC 6946 X-1 and NGC 1313 X-1 fall on the power law $F_{rms} \propto f_q^{-\gamma}$ with a single exponent $\gamma \approx 0.3$. The source M 82 X-1 differs from them and has the exponent $\gamma \approx 0.17$.

Figures 14b and 14c show the normalized counting rate (reduced to the same distance) and the spectral hardness of the source as a function of the QPO frequency. The objects differ significantly from each other in terms of hardness. The figure shows that the harder the source is, the higher are frequencies at which QPO is observed: IC 342 X-1 has a very hard spectrum (one of the hardest ULXs, besides ultraluminous pulsars) and the highest QPO frequency among the five objects studied. The softest NGC 6946 X-1 and NGC 5408 X-1 show frequencies much lower. However, if we talk about each specific source, then variations in its QPO frequency between observations do not affect the hardness. The only exception is NGC 6946 X-1, in which we found a weak positive correlation between frequency and hardness. We also found a positive correlation between frequency and luminosity (Fig. 14b).

It is worth noting that QPO in the studied objects appears not in all the observations (for example, in IC 342 X-1, the QPO peak is found only in one). Therefore, in Fig. 14d we plotted the relative variability vs. counting rate for all observations regardless of the presence of QPO. It can be seen that observations with and without QPO form a single sequence, and at times when QPO are absent, the sources are brighter and less variable. It can be supposed that each source has a certain luminosity threshold, above which the mechanism producing QPO and FTN breaks down and the variability disappears. It is possible that these variations of the shape of the power spectra are related to fluctuations in the accretion rate and, accordingly, the strength of the outflowing wind.

Using the above relationships, we compared the masses of the black holes of these five objects. Expressing the masses of black holes in units of NGC 5408 X-1 (1.0), we obtained masses of 0.9, 9.5, 1.6 and 1.8 for NGC 6946 X-1, M 82 X-1, NGC 1313 X-1 and IC 342 X-1, respectively. In the case of M 82 X-1, it turned out that the black hole should be about 10 times more massive than in other objects. Here we assumed that the maximum X-ray luminosity depends only on the mass of the black hole and the accretion rate, but nevertheless more accurate estimates should also take into account other parameters, namely the collimation of radiation and the inclination of the disk.

In other papers, we investigated the longest observations of SS 433 (Atapin and Fabrika, 2016) and ULXs (Atapin and Fabrika, 2017; Fabrika et al., 2018) (the objects NGC 5408 X-1 and NGC 6946 X1, for each of them there were several XMM-Newton observations with a length of about $10^5$ s). This allowed us to extend the power spectra to frequencies of $10^{-5}$ Hz in the case of the ULXs and to $10^{-6}$ Hz for SS 433 (Fig. 15). It can be seen that both power spectra have a similar shape: each of them has a flat section, but its length is limited to two or three orders of magnitude in frequency. At lower and higher frequencies, the power spectra have a power-law form with exponents $1.5-2$.

We assume that, both the presence of a flat section in the power spectra of a supercritical disk and the (anti)correlation between QPO frequency and fractional variability can be explained within the framework of the idea of Lyubarskii (1997). According to this model, random viscosity fluctuations at different disk radii have a characteristic time scale of the order of viscosity time $t_{visc}(R) = [\alpha(H/R)^2 \Omega_K(R)]^{-1}$, where $\alpha$ is the viscosity parameter (Shakura and Sunyaev, 1973), $\Omega_K$ is the

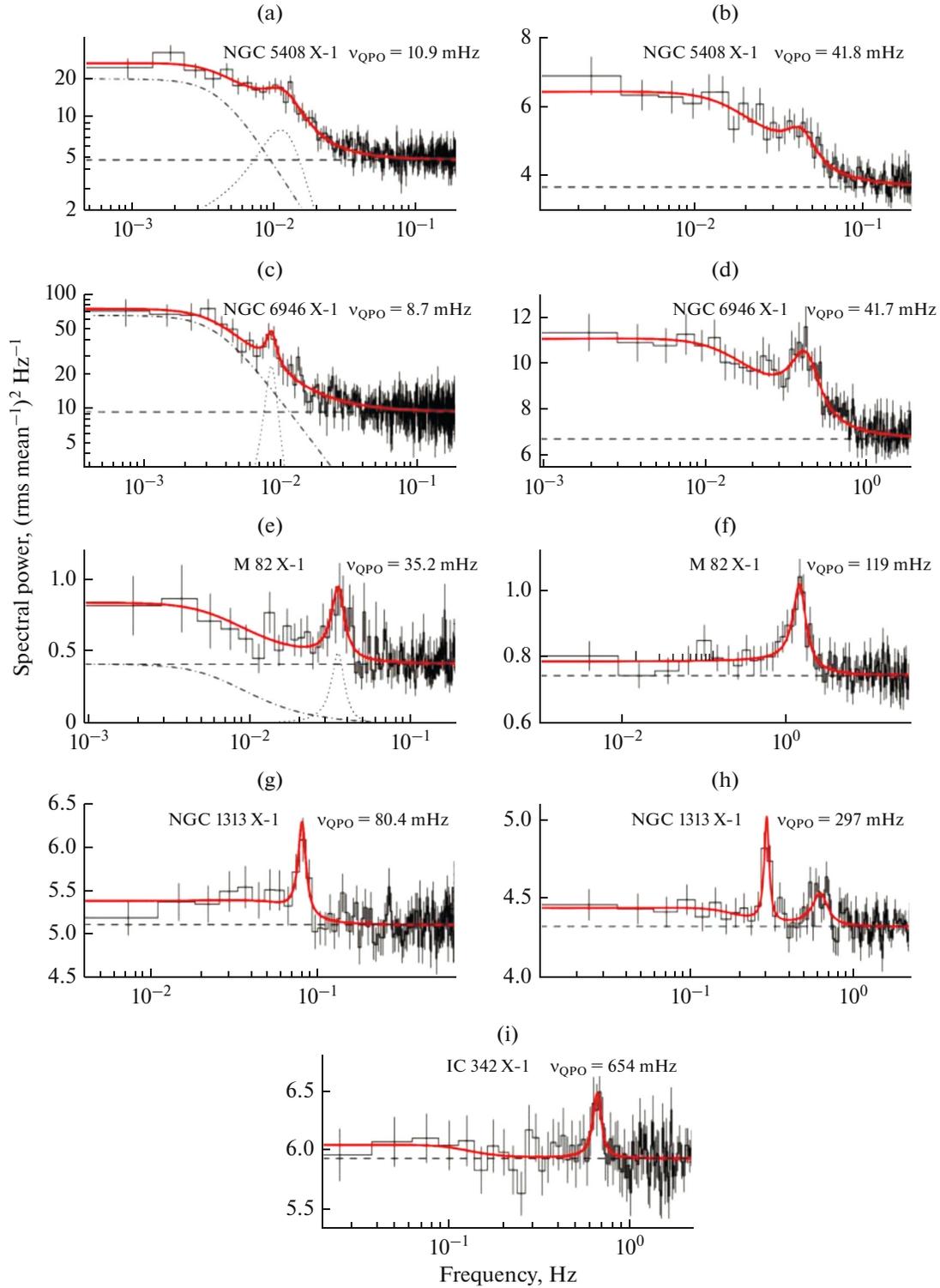

**Fig. 13.** Power spectra of the objects NGC 5408 X-1, NGC 6946 X-1, M 82 X-1, NGC 1313 X-1 and IC 342 X-1 in the energy range 1−10 keV. The power spectra with the lowest and highest QPO frequency are shown on the left and right. The solid line is the best model; the dotted and dash-dotted lines show its components: the Lorentzian for approximating the QPO peak, the broken power law for describing flat-topped noise. The dashed line indicates the level of Poisson noise. It can be seen that in the power spectra with a lower QPO frequency, the level of the flat section is higher.

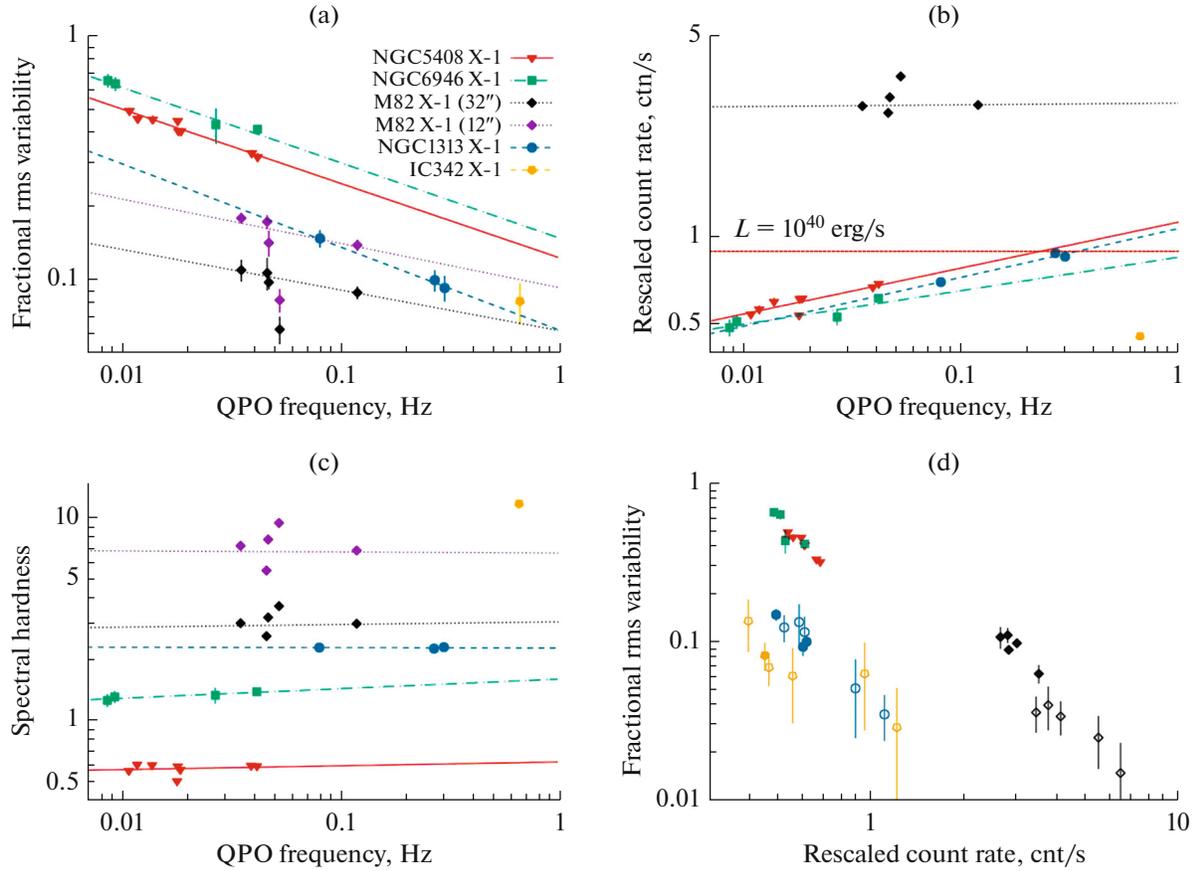

**Fig. 14.** Fast variability of five ultraluminous X-ray sources. (a) The fractional rms variability $F_{rms}$ in the energy range of $1-10$ keV as a function of the QPO frequency. (b) Count rate in the range of $1-10$ keV, reduced to the same distance (5.32 Mpc, the galaxy NGC 5408), vs. the QPO frequency. The red dashed line shows the count rate corresponding to the luminosity $10^{40}$ erg s$^{-1}$. (c) Spectral hardness (ratio of fluxes in the range $1-10$ and $0.3-1$ keV) vs. QPO frequency. (d) Relative variability depending on the reduced count rate. Open symbols indicate observations in which QPOs are not detected. For the object M 82 X-1, the data obtained from two apertures are presented: 32″ (as for the other objects) and 12″. The smaller aperture was used to reduce the contribution of the ultraluminous pulsar M 82 X-2 located 6″ away from it.

Kepler frequency. The viscosity time decreases as we approach the black hole, so as matter passes through the disk, the slow large-scale fluctuations that occur in the peripheral regions of the disk are superimposed with faster ones. Since the viscosity time decreases smoothly in a standard disk, the power spectrum turns out to be power-law (Lyubarskii, 1997).

In a supercritical disk, the situation is different. Inside the spherization radius, the thickness-to-radius ratio is $H/R \sim 0.7$ (Lipunova, 1999). Beyond the spherization radius, the disk has a structure similar to the standard disk—$H/R \sim 0.03-0.1$ (Shakura and Sunyaev, 1973). As a result, when the matter crosses the spherization radius, the viscosity time should decrease sharply. We believe that in this case, the spherization radius plays the role of a trigger that controls the flow of matter into the supercritical region of the disk, and if the viscosity on it changes randomly

(white noise), as was assumed for all other radii of the disk, then a flat section should appear in the power spectrum.

Figure 15a shows what the power spectrum should look like within this model. It has a flat section and two breaks: a break at the frequency $f_b$, which is clearly observed in ULXs (Fig. 13) and SS 433 (Fig. 12), as well as a low-frequency break at $f_{b,\,low}$. In our opinion, the appearance of the flat section is associated with fluctuations directly on the spherization radius $R_{sp}$, and the break frequency $f_b$ is determined by the viscosity time at $R_{sp}$. The power-law section at frequencies above the break is formed in the supercritical region of the disk, where the viscosity time monotonically decreases from $t_{visc}(R_{sp})$ to $t_{visc}(R_{in})$. The low frequency region should correspond to the peripheral areas of the disk. We assume that at low frequencies the power spectrum should again become power-law, since in the region $R \gg R_{\text{sp}}$ the disk structure should be similar to the standard one. There should be a break at $f_{b,\text{low}}$, after which the flat section is replaced by a power-law one.

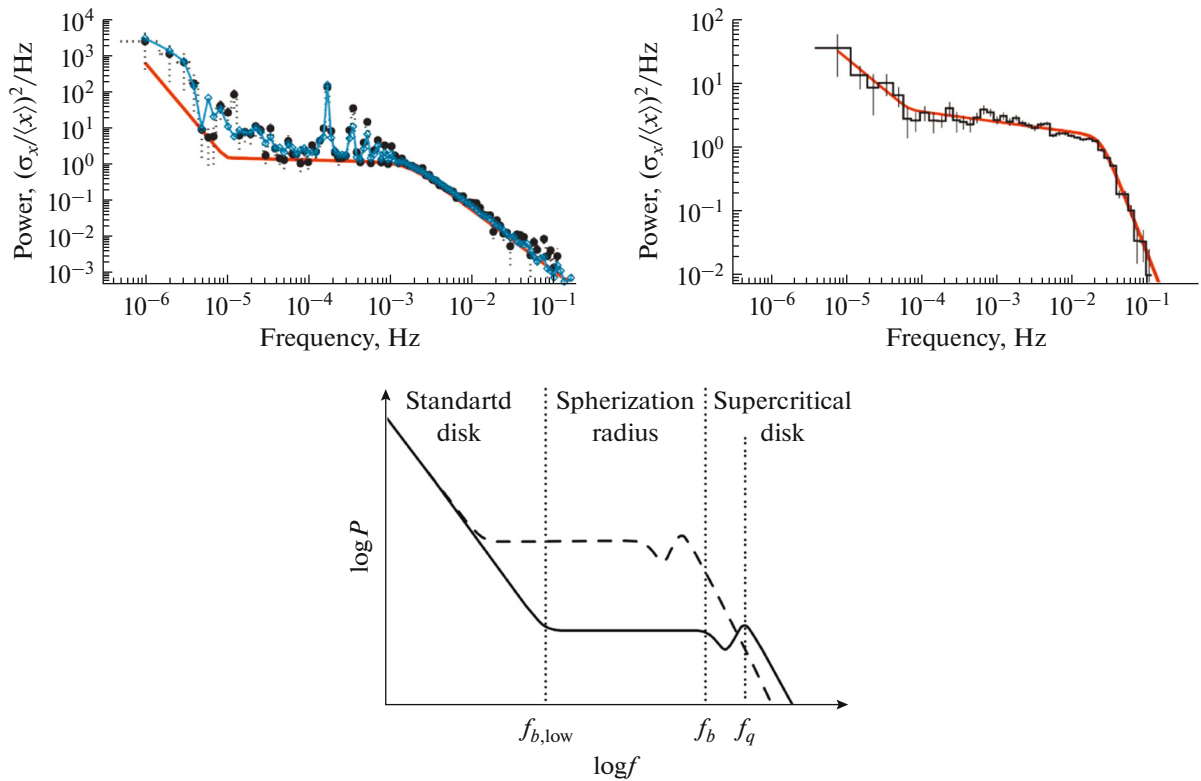

**Fig. 15.** (a) The power spectrum of SS 433, constructed from the data of a very long, of the order of $10^6$ s, ASCA observation. The black dots show the observed power spectrum. Since the observation had gaps, false peaks and other artifacts appear in the power spectrum. The red solid line shows the original model, the blue—Monte Carlo modeling of distortions that appear in the original model as a result of adding the same gaps as in the observations. It can be seen that the model describes the emerging peaks well. (b) The power spectrum of the ultraluminous X-ray source NGC 5408 X-1, obtained from XMM-Newton data. It can be seen that both power spectra are very similar, and each of them has a flat section with a length of $2-2.5$ orders of magnitude in frequency. (c) The sketch of the power spectrum of fast variability occurring in a supercritical accretion disk. The following frequencies are marked: $f_q$—the QPO frequency (Fig. 13), $f_b$ and $f_{b,\text{low}}$—high- and low-frequency breaks. All three characteristic frequencies depend on the accretion rate, which determines the size of the supercritical zone of the disk (spherization radius) (Atapin et al., 2015, 2019). A flat section occurs due to fluctuations in viscosity at the spherization radius. As the accretion rate increases, all frequencies shift to the left in the figure (decrease), and the level of the flat section increases (shown by the dashed line).

## 3. STUDY OF ULTRALUMINOUS X-RAY SOURCES IN THE OPTICAL RANGE

Optical (as well as infrared and ultraviolet) observations of ULXs can provide a lot of important information beyond what can be obtained from the analysis of X-ray data. ULXs, due to the fact that they are close binary systems, in the optical range should look like point star-like sources. Identifying ULXs with such sources (the so-called *optical counterparts*) can allow, for example, to construct curves of the radial velocities of companion stars and, thus, put restrictions on the mass of compact objects, which still remains one of the most important issues. The answer to this crucial question has not yet been received. Such measurements are rarely carried out, because in many cases the radiation of the donor star is dominated by the radiation of a supercritical disk (see below), the outer regions of which emit in the optical range. In this case, however, we are able to obtain information about the accretion flow and wind. Finally, optical studies of the ULX environment (stellar population and nebulae) can shed light on the history of the evolution of binary systems.

### 3.1. Main Characteristics of the ULX Environment

The study of the environment of ultraluminous X-ray sources showed the presence of bubble nebulae around many of them ranging in size from several tens to hundreds of parsecs (Cseh et al., 2012; Grise et al., 2012; Pakull and Mirioni, 2003). The shapes of these nebulae, as well as changes in the radial velocities in them with an amplitude of about 100 km s$^{-1}$ (Fabrika et al., 2006; Lehmann et al., 2005; Pakull et al., 2006) indicate the existence of an additional energy source that dynamically perturbs the interstellar medium. As was shown in Abolmasov et al. (2007), this energy source can be shock waves that occur when relativistic jets or wind collide with surrounding medium. Bright low-excitation lines, such as [O I] 6300, 6364 Å and [N II] 5200 Å, as well as high line intensity ratios [S II] 6717, 6731 Å/Hα > 0.3 and [N II] 6548, 6583 Å/Hα > 0.5, visible in the nebulae spectra around some ULXs (for example, IC 342 X-1, M 51 X-1), imply electron impact ionization, which in most cases can be explained by the presence of shock waves with velocities of 20−100 km s$^{-1}$ (Abolmasov et al., 2007). At the same time, the total energy of shock waves estimated from the luminosity of the Hβ line is comparable to the X-ray luminosities of ULXs, which is completely consistent with the expected wind power at supercritical accretion.

Another source of nebulae gas excitation—much more effective than X-ray radiation—can be photoionization by extreme ultraviolet (EUV), which can account for a significant portion of the radiation of supercritical disks (Abolmasov et al., 2007, 2008; Vinokurov et al., 2013). At the same time, the collimation of ultraviolet radiation is not expected to be very high (a factor of a few (Ohsuga et al., 2005)), which is sufficient to obtain both the observed luminosities and the observed shapes of nebulae. The relatively bright lines He II 4686 Å and [Fe III] and very bright lines [O III] (the ratio [O III/H β > 3]), which are observed in the spectra of nebulae surrounding NGC 6946 ULX-1, NGC 5204 X-1 and other objects, are evidence of the presence of a powerful photoionization source in EUV. It is important to note that although a luminous (up to $10^{40}$ erg s$^{-1}$) ultraviolet source is required to explain the spectra of many (but not all) nebulae, the signs of shock excitation are characteristic of all nebulae surrounding ULXs (Abolmasov et al., 2007).

Many ultraluminous X-ray sources (mainly in starburst galaxies) are located near dense compact star clusters or super star clusters (SSCs), which suggests their physical connection. The most clearly the presence of this connection was confirmed in the work Poutanen et al. (2013), where the study of ULXs in Antenna galaxies (NGC 4038/NGC 4039) was conducted. At the same time, it was shown that most of the luminous X-ray sources are located nearby (at distances up to 290 pc), but not inside star clusters. The clusters turned out to be very young, their age is less than 6 Myr. This allowed us to conclude that the masses of ULX progenitors exceed 30 $M_\odot$, and in some cases should reach up to 100 $M_\odot$. The results obtained are consistent with the idea that most ULXs are massive X-ray binaries that were ejected during the formation of star clusters as a result of multiple star collisions (Portegies Zwart et al., 2004). Similar results were obtained for another galaxy with strong star-forming—NGC 3256. In both studies, the age of clusters was determined by comparison of spectroscopic and photometric data with the results of modeling the spectra, fluxes of emission lines He II 4676 Å and C IV 5808 Å and spectral energy distributions (SEDs) of star clusters using the code Starburst99 (Leitherer et al., 1999).

Nevertheless, bright ULXs are not always associated with young clusters, even in galaxies with a very high rate of star formation, located in non-star-forming regions (see, for example, Soria et al. (2012)). This fact may support the existence of two populations of ultraluminous X-ray sources with significantly different ages in such galaxies. In spiral and dwarf galaxies with moderate star formation, ULXs can be found near "loose" clusters or OB associations with masses of several thousand solar masses and ages of 10−20 Myr (Feng and Kaaret, 2008; Grisé et al., 2008, 2011). In elliptical galaxies, there is a deficit of bright ULXs ($L_X \gtrsim 2-3 \times 10^{39}$ erg s$^{-1}$), which is consistent with a gap below $10^{39}$ erg s$^{-1}$ in the X-ray luminosity function of elliptical galaxies not affected by recent star formation (Kim and Fabbiano, 2010; Sarazin et al., 2000). Among the few ULXs with luminosities in the bright state $L_X = 4-5 \times 10^{39}$ erg s$^{-1}$, located in elliptical galaxies, we can mention a transient ultraluminous X-ray source M 86 TULX-1 (van Haaften et al., 2019), ULXs in globular clusters, for example, in NGC 1399 (Feng and Kaaret, 2006) and NGC 4472 (Maccarone et al., 2007).

Information on the ages of the stellar population surrounding ULXs makes it possible to find restrictions not only on the masses of the progenitors of compact objects (as in the work Poutanen et al. (2013)), but also the upper limits on the masses of donor stars. Using this method, many studies have shown that donors in a significant part of ULXs have masses of no more than 10−15 $M_\odot$ (for example, Feng and Kaaret (2008); Grisé et al. (2008)). Nevertheless, much more information about donor stars is provided by photometry and spectroscopy of ULX optical counterparts. Here, we just note that the estimates obtained at least do not contradict the modern results of the population synthesis of X-ray binaries (Wiktorowicz et al., 2017). In these calculations, the authors found that a typical ULX with a black hole should contain a donor star with a mass of 6 $M_\odot$ on the Main Sequence, and red giants with a mass of 1 $M_\odot$ should be companions of neutron stars.

### 3.2. Optical Counterparts

Catalogs of ultraluminous X-ray sources (Earnshaw et al., 2019; Liu, 2011; Walton et al., 2011) include about 400 candidates, but only a few dozen objects are identified and relatively well studied in the optical range (Gladstone et al., 2013; Ptak et al., 2006). In most cases, optical identifications are not unambiguous: several optical sources fall within the circle of X-ray source coordinate errors. Only about 20 objects have reliable optical counterparts for which spectral energy distributions over a wide range of wavelengths are known (see, for example, Tao et al. (2011); Vinokurov et al. (2018)).

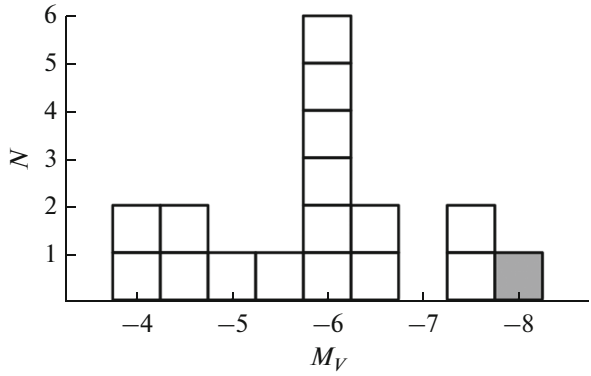

**Fig. 16.** Absorption-corrected absolute stellar magnitudes of ultraluminous X-ray sources and SS 433 (shown in gray). In order of decreasing brightness of the object: SS 433, NGC 6946 ULX-1, NGC 7793 P13 (ULXP), NGC 4559 X7, NGC 5408 X-1, NGC 5204 X-1, NGC 4395 X-1, M 81 ULS-1, Holmberg II X-1, IC 342 X-1, Holmberg IX X-1, NGC 4559 X-10, NGC 1313 X-2 (ULXP), NGC 5474 X-1, NGC 1313 X-1, M 66 X-1 and M 81 X-6 (Fabrika, 2016; Vinokurov et al., 2018).

Most of the optical counterparts of ultraluminous X-ray sources were found from data of the Hubble Space Telescope (HST) (Avdan et al., 2019; Earnshaw and Roberts, 2017; Gladstone et al., 2013; Kaaret et al., 2010; Liu et al., 2007; Ptak et al., 2006; Ramsey et al., 2006; Roberts et al., 2008; Soria et al., 2005; Tao et al., 2011; Yang et al., 2011). In some cases, optical counterparts of ULXs were detected as a result of studies performed on ground-based telescopes (for example, NGC 7793 P13 (Motch et al., 2011, 2014)). However, such identifications are rare, since most ultraluminous X-ray sources are located in very crowded stellar fields, and their unambiguous identification is possible only on the basis of HST data.

In the optical range, ultraluminous X-ray sources are faint objects, the brightest of which have an apparent magnitude of about $20^m$. The apparent magnitudes of most ULX optical counterparts are in the range $m_V = 21-24$ (Gladstone et al., 2013; Tao et al., 2011), but there are also much fainter objects with $m_V = 25-26$ (see, for example, Avdan et al. (2016)). The brightness variability of ULX optical counterparts rarely exceeds $\Delta V = 0.1-0.2$ stellar magnitude (Tao et al., 2011), although those objects whose X-ray luminosity varies by a factor of hundred or more often show strong changes (up to several stellar magnitudes) in the optical range as well (Motch et al., 2014; Soria et al., 2012; van Haaften et al., 2019).

For ULXs, the ratio of X-ray and optical fluxes $F_X/F_{opt}$ ranges from several hundred to several thou-sand. At that, the boundaries of this range is quite insensitive to the measurement methods, which may differ both in the set of optical filters used and in the X-ray ranges (see Avdan et al. (2016); Soria et al. (2012); Tao et al. (2011)). Similarly high values of $F_X/F_{opt}$ are observed in low-mass Xray binaries (LMXBs), while high-mass X-ray binaries (HMXBs) show lower ratios. This similarity between ULXs and LMXBs suggests that the contribution of the donor star to the optical emission of ULXs may also be very small. However, in practice, the picture is less clear (see below).

The high $F_X/F_{opt}$ ratio observed in ultraluminous X-ray sources is of great practical importance: it allows to distinguish ULXs from active galactic nuclei (AGNs). The fact is that AGNs are often projected onto mimicking the X-ray sources-members of these galaxies, so using only X-ray data, it is not always possible to distinguish them. This leads to a significant proportion of background AGNs that were mistakenly included in the ULX candidate catalogs. In the work Vinokurov et al. (2018), we have refined the criterion that can be used to separate two classes of objects: $F_X/F_{opt} > 100-200$. Indeed, in the vast majority of AGNs, the ratio of X-ray to optical luminosity does not exceed 10 (Aird et al., 2010), and only in some cases in galaxies with very high internal absorption, this ratio reaches a value of the order of 100 (Della Ceca et al., 2015).

The distribution of ultraluminous X-ray sources by absolute stellar magnitudes in the V band (Fig. 16) shows an obvious peak at $M_V = -6$, and the decrease in the number of sources with a decrease in their brightness seems to be physical, not selective (Tao et al., 2011; Vinokurov et al., 2018). Common properties of bright (with absolute magnitudes $M_V \lesssim -6$) optical counterparts of ULXs are their blue power spectral energy distributions $F_\nu \propto \nu^\alpha$ with the exponent $\alpha$ in the range 1–2 (Tao et al., 2011; Vinokurov et al., 2013). The SEDs of such shapes are in good agreement with what is expected in the case of the dominance in the optical range of hot winds of supercritical disks with a relatively small contribution of donor stars (Fabrika et al., 2015). One of the few exceptions is the ultraluminous pulsar NGC 7793 P13 ($M_V \approx -7.5$), in which the class B9Ia donor dominates even when the object is in the bright state (Motch et al., 2014).

Most of the weaker in the optical range ultraluminous X-ray sources (with $M_V > -5.^m5$) are objects with relatively "cold" energy distributions corresponding to supergiants of classes A—G (Avdan et al., 2016, 2019), which may indicate the predominance of the contribution of donor stars over the contribution of hot winds of supercritical disks (Vinokurov et al., 2018). The reason for small contribution of wind radiation may be lower accretion rates in these objects. As shown in our paper Fabrika et al. (2015), the optical luminosity in the supercritical accretion regime will depend radically on the rate of gas outflow, which in turn is comparable to the initial accretion rate $\dot{M}_0$ (see Introduction). This dependence can be deduced on the basis of simple relations. The expected value of the wind velocity at the spherization radius of the supercritical disk is close to the virial $V(R_{\rm sp}) \propto M^{1/2} R_{\rm sp}^{-1/2}$ (Shakura and Sunyaev, 1973). In this case, the radius of the wind photosphere can be estimated as $R_{\rm ph} \propto \dot{M}_0 V^{-1} \propto \dot{M}_0^{3/2} M^{-1/2}$ (Fabrika, 2004). The bolometric luminosity of supercritical disks with a wind photosphere is determined by the relation $L \propto M \propto R_{\rm ph}^2 T_{\rm ph}^4$. By combining these two relations, one can estimate the photospheric temperature $T_{\rm ph} \propto \dot{M}_0^{-3/4} M^{1/2}$ (Fabrika et al., 2015). Since at the expected high temperatures of the wind photosphere (of the order of several tens of thousands of degrees (Fabrika et al., 2015)), the optical radiation of the wind should fall on the Rayleigh–Jeans region, hence $L_V \propto R_{\rm ph}^2 T_{\rm ph} \propto \dot{M}_0^{9/4} M^{-1/2}$. In the case of a constant gas velocity in the wind, the dependence is slightly weaker: $L_V \propto \dot{M}_0^{3/2} M^{1/4}$ (Fabrika et al., 2015).

The size of the wind photosphere and its luminosity in the optical range are determined mainly by the rate of outflow of matter in the wind, and the higher it is, the brighter the object will be. Detailed modeling of the ULX optical spectra allowed us to estimate the outflow rates of some of the brightest sources (Kostenkov et al., 2020, see Section 3.3 in this paper): they turned out to be of the order $\dot{M} = 10^{-5} - 10^{-4}\,M_\odot\,{\rm yr}^{-1}$. Numerical estimates of the photosphere radius can be obtained from the relation written in the previous paragraph, which, taking into account the coefficients for the case of a constant wind velocity $V \approx 1000\,{\rm km\,s}^{-1}$ (Section 3.3) will take the form $R_{\rm ph} = \kappa \dot{M}/(\Omega V)$, where $\Omega$ is the solid angle of the wind, $\kappa$ is the opacity. The simplest case, which simultaneously gives lower estimates for $R_{\rm ph}$, assumes complete ionization of the wind and its spherical symmetry (then $\kappa$ is the Thomson opacity, $\Omega = 4\pi$). In this approximation, the radius of the wind photosphere is $2 \times 10^{11} - 2 \times 10^{12}$ cm. At temperatures of about 30000 K (Kostenkov et al., 2020), the lower estimate of the bolometric luminosity of the wind photosphere will exceed 5000 $L_\odot$.

Detecting objects with $M_V > -5^{\rm m}.5$ can provide clues as to which classes of stars act as donors in ULXs. The fact that many such systems contain relatively cool supergiants (up to class M) is also confirmed by studies of the infrared spectra of ULXs (Heida et al., 2016, 2015; López et al., 2020). Nevertheless, these conclusions are based on the study of a rather small number of objects (about 20 optical counterparts in the entire $M_V$ range), and their confirmation requires a significant increase in the sample of the studied ULXs.

In most cases, the constraints on the spectral types and luminosity classes of ULX donor stars obtained from the analysis of optical data provide only some hints on of the accretor type. The situation is different when an independent method helps to determine the orbital period of the system. Then, with the restrictions on the mass of the companion star obtained from optical data, it is possible to determine the mass of a compact object fairly reliably. A good example of this is the eclipsing system CXOM51 J132946.1+471042 (see above), in which a massive donor is paired with a non-pulsating neutron star (Wang et al., 2018). Using HST data, it was possible to detect the optical counterpart of this ULX and obtain restrictions on the donor mass of 20–35 $M_\odot$. Combining these data with information about the orbital period found from eclipses, the authors showed that the mass of the donor must exceed the mass of the relativistic component of the system by at least 18 times. This, in turn, allowed to estimate the mass of the accretor, which turned out to correspond to the mass of the neutron star. However, even more wide possibilities for determining the masses of both components of the binary system are provided by the detection of donor lines in the spectrum of the optical counterpart, although this can be done quite rarely (see Section 3.4).

### 3.3. Optical Spectra

Spectroscopy of ultraluminous X-ray sources in the optical range began as soon as the first unambiguous optical identifications were performed (see, for example, Grisé et al., 2009; Pakull et al., 2006). Although about fifteen years have passed since the first results were obtained, so far optical (as well as UV (Bregman et al., 2012) and IR (Heida et al., 2016, 2015; López et al., 2020)) spectra available only for a very small part of ULXs. In the optical range, ten objects are most well studied: NGC 1313 X-2 (Grisé et al., 2009; Pakull et al., 2006; Roberts et al., 2011); NGC 5408 X-1 (Cseh et al., 2011, 2013); NGC 7793 P13 (Motch et al., 2014); NGC 4559 X-7, NGC 5204 X-1, Holmberg IX X-1, Holmberg II X-1 (Fabrika et al., 2015); NGC 4395 ULX-1 (Vinokurov et al., 2018); NGC 300 ULX1 (Heida et al., 2019; Villar et al., 2016); UGC 6456 ULX (Vinokurov et al., 2020). All objects in the bright state have X-ray luminosities above $3 \times 10^{39}$ erg s$^{-1}$, characteristic of bona fide ULXs, and three objects: NGC 1313 X-2, NGC 7793 P13, and NGC 300 ULX-1 are known ultraluminous X-ray pulsars. The spectra were obtained on large ground-based telescopes: Subaru of the Japanese National Astronomical Observatory, VLT (Very Large Telescope) of the European Southern Observatory, on the 6-m BTA telescope of SAO RAS. The small number of optical counterparts stud-ied is due to the difficulty of spectral observations because of the faintness of ULXs in the optical range (the $V$-band magnitudes from $20^m$ to $26^m$), and also because of the narrow star fields in which most of them are located. The latter puts serious limitations on image quality (usually significantly better than 1″ is required) for ground based observations.

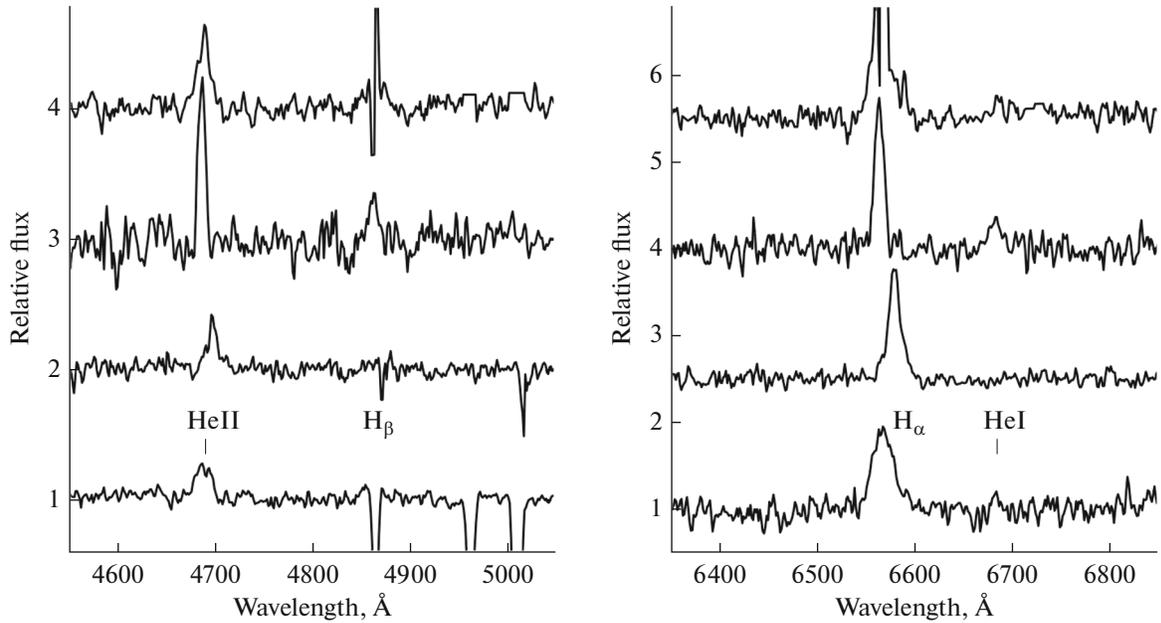

**Fig. 17.** Normalized spectra of Holmberg II X-1, Holmberg IX X-1, NGC 4559 X-7 and NGC 5204 X-1 (from top to bottom) in the blue (a) and red (b) spectral ranges obtained in February 2011 by the Subaru telescope (Hawaii). The spectra are not reduced to zero radial velocity. The brightest wide emission lines are He II 4686 Å, Hβ, Hα, and He I 6678 Å. The narrow emission lines of the nebulae Hβ and [O III] 4959, 5007 A in the cases of NGC 4559 X-7 and NGC 5204 X-1 were oversubtracted during the extraction of the spectra of objects (see more details in Fabrika et al. (2015)), however, the wide wings of the line Hβ are clearly visible.

The optical spectra of ULXs are characterized by the presence of wide ($FWHM \approx 300{-}1600$ km s$^{-1}$) emission lines (Fabrika et al., 2015; Roberts et al., 2011) (Fig. 17). The most frequently observed lines are HeII 4686 Å and the lines of the Balmer hydrogen series. The line He II 4686 Å was detected in the spectra of all the studied objects, while wide hydrogen lines were detected only in eight ULXs. The exceptions are NGC 1313 X-2 and NGC 4395 ULX-1. In the works devoted to spectral studies of NGC 1313 X-2 (Grise et al., 2009; Pakull et al., 2006; Roberts et al., 2011), the detection of other broad lines, except for He II, is not reported. Similarly, in the spectra of NGC 4395 ULX-1, the He II line is the only one in which the wide component was detected (Vinokurov et al., 2018), but in this case the possibility of detecting weaker wide lines is limited by the low $S/N$ ratio in the spectra.

Neutral helium emissions are slightly less frequently observed (four objects from the work Fabrika et al. (2015), UGC 6456 ULX (Vinokurov et al., 2020), NGC 300 ULX-1 (Villar et al., 2016)), which may also be related to the threshold for detecting weak lines in the spectra. Note that the search for broad components of emission lines is often complicated by the presence of bright, but narrow, emission lines of nebulae surrounding many ULXs (Pakull and Mirioni, 2003).

The ultraluminous X-ray pulsars NGC 7793 P13 and NGC 300 ULX-1 show spectra richer in different lines. In addition to the broad components of the Balmer series and He II 4686 Å lines, both objects have bright emissions of heavier elements. In the spectrum of NGC 7793 P13, the blend C III/N III 4640–4650 Å is clearly visible (Motch et al., 2014) (although hints of its presence are also present in the spectra of the three ULXs from the work Fabrika et al. (2015)). The spectrum of NGC 300 ULX-1 is full of allowed and forbidden lines of iron, calcium and other elements (Villar et al., 2016). In addition, donor-owned absorption lines were detected in the spectra of NGC 7793 P13 and NGC 300 ULX-1 (Heida et al., 2019; Motch et al., 2014).

In Fabrika et al. (2015), using the spectra of five ULXs with well-distinguishable wide hydrogen and helium emissions, we measured the ratios of their equivalent widths:

$$EW\,(\mathrm{He\,II})/EW\,(\mathrm{H\beta}) \approx 2.2,$$

$$EW\,(\mathrm{He\,II})/EW\,(\mathrm{H\alpha}) \approx 0.36,$$

$$EW\,(\mathrm{He\,II})/EW\,(\mathrm{He\,I\,5876}) \gtrsim 3.6.$$

Based on these estimates, a conclusion was made about the high temperature of the gas (several tens of thousands of degrees) in which these lines formed. At the same time, the ratio of hydrogen and helium was estimated to be close to the solar content, since the Pickering series lines are weak: $EW(\text{He II } 5411\,\text{Å})/EW(\text{H}\beta) \lesssim 0.27$.

To date, the most studied is the behavior of the He II 4686 Å line. Significant variability of its profile width, equivalent width, and radial velocity was found on time scales from one day to months. *FWHM* of the line can vary from 13% (NGC 5408 X-1 (Cseh et al., 2013)) to 3 times (NGC 4559 X-7 and NGC 5204 X-1 (Fabrika et al., 2015)) and even slightly more in the case of NGC 1313 X-2 (Roberts et al., 2011). The variability of the radial velocity in the range of 100–400 km s$^{-1}$ was found in seven of the ten objects listed above. Based on the analysis of series of observations for one of them, NGC 7793 P13, an orbital period of 63–65 days was determined (Motch et al., 2014). For the objects NGC1313 X-2, Holmberg IX X-1, NGC 5408 X-1, no statistically significant periodic signal was detected, although the possibility of short periods of the order or less than one day is not excluded (Cseh et al., 2013; Roberts et al., 2011). For the remaining three objects (NGC 4559 X-7, NGC 5204 X-1 and UGC 6456 ULX), the amount of available data is still insufficient to search for periodicity. The behavior of the lines of the Balmer series and the lines of heavy elements rarely found in the ULX spectra is less studied, but in general the variability time-scales and amplitudes of these lines are comparable to those of He II. The deepest analysis of the behavior of some lines of the Balmer series was carried out in the works Fabrika et al. (2015); Motch et al. (2011).

In early studies, it was believed that the region of formation of wide emission lines is a thin accretion disk heated by X-ray radiation. It is known that the emission He II is formed in hotter regions than the lines of the Balmer series. In disks, such regions are located closer to the accretor and have higher Keplerian speeds, therefore, the He II line should be significantly wider than the hydrogen lines. This pattern is observed in the spectra of Galactic X-ray transients (for example, GX 339-4 (Rahoui et al., 2014; Soria et al., 1999), V404 Cyg (Casares et al., 1991; Gotthelf et al., 1992), GRO J165540 (Hunstead et al., 1997; Soria et al., 1998)). However, in the case of ULXs, the hydrogen lines are on average 30% wider than the He II line (Fig. 18). This fact was explained by us in Fabrika et al. (2015) based on the idea that the lines are formed in radiatively accelerated winds, which have to be present in the case of super-Eddington accretion. Indeed, the wind accelerates gradually, so its colder regions, which are further away from the accretor, have a higher velocity than those that are close and hot. Schematically, the regions of line formation are shown in Fig. 4. This effect is well known and is observed in the spectra of stars with a wind photosphere: luminous blue variables (LBV) and late Wolf–Rayet stars of the nitrogen sequence with hydrogen lines (WNLh). The spectra of these stars are very similar to those of ultraluminous X-ray sources. However, the observed optical spectra of ULXs cannot be formed in the donor winds, even if the donors are stars of the aforementioned types or Wolf–Rayet stars of other subclasses. The presence of strong hydrogen lines in the ULX spectra contradicts the latter. In addition, in the case of Wolf–Rayet donors, accretion must be wind-fed, and it is able to provide the ULX luminosities only in the case of sufficiently short orbital periods, such as those of M101 ULX-1 and CG X-1 (Liu et al., 2013; Qiu et al., 2019). A more powerful argument against the idea of the stellar origin of the ULX spectra is the strong variability of the width of the emission lines at times less than one day, which is not typical for stars, since the terminal wind velocity is determined mainly by gravity on the star surface and its luminosity, and therefore can only change over large time intervals.

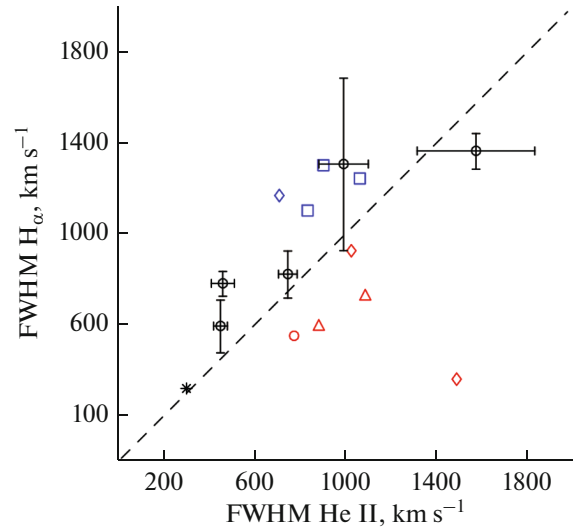

**Fig. 18.** The widths (*FWHM*) of the emission lines He II and Hα of the ultraluminous X-ray sources Holmberg IX X1, NGC 4559 X-7, NGC 5408 X-1, Holmberg II X-1, NGC 5204 X-1 (circles with error bars located from left to right in the specified order). Due to the strong variability of the lines, measurements of the widths of He II and Hα were carried out using the Subaru telescope data obtained during one single night (Fabrika et al., 2015). The line widths of NGC 5408 X-1 are taken from the VLT archive data (Cseh et al., 2013). The values of the error bars takes into account the systematic error associated with the uncertainty of subtracting the nebula contribution. For comparison, the ratios of the widths of these lines are given for the transitional stars WR 22, WR 24 and WR 25 (all three are of the type WN 6ha (Walborn and Fitzpatrick, 2000), blue squares), V 532 (LBV in the hot state in the galaxy M 33 (Sholukhova et al., 2011), indicated by an asterisk) and SS 433 (Grandi et al. (1982); Kubota et al. (2010), blue rhombus, measurements are made on the spectra in the same precession phase). Red symbols show the position of X-ray transients with black holes GX 339-4 (Rahoui et al. (2014); Soria et al. (1999) circle), GRO J1655-40 (Casares et al. (1991); Gotthelf et al. (1992), rhombuses) and V 404 Cyg (Casares et al. (1991); Gotthelf et al. (1992), triangles). For the last two objects, there are two measurements each. Four of the five ULXs have the lines He II wider than Hα. The same pattern is observed for the stars WNLh and LBV, which have powerful winds. In contrast to these objects, all transients are located below the dashed line that marks the place of equal widths of the lines He II and Hα.

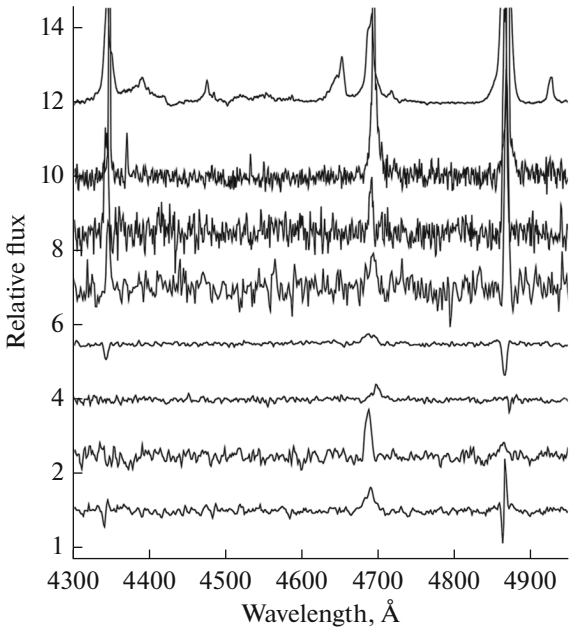

**Fig. 19.** Normalized spectra of seven ULXs and SS 433. From top to bottom: SS 433 (1), NGC 5408 X-1 (2), NGC 4395 ULX-1 (3), NGC 1313 X-2 (2), NGC 5204 X-1 (1), NGC 4559 X-7 (1), Holmberg IX X-1 (1) and Holmberg II X-1 (1). The numbers in parentheses correspond to the telescopes on which the observations were made: *1*—Subaru, *2*—VLT, *3*—BTA. To the right of the line Hγ in the spectrum of SS 433, the relativistic line H$^-$β is visible. As in Fig. 17, there are traces of oversubtraction of the narrow component of the line Hβ, which is formed in the H II regions surrounding ULXs.

SS 433 also has a spectrum that has a wind origin, with the brightest spectral features being the same as observed in the ULX spectra. The exception is "moving" jet lines (for more information, see Section 3.5), which ULXs do not exhibit. The normalized spectra of seven ULXs in comparison with the spectrum of SS 433 are shown in Fig. 19. As a result of comparing the observational properties of ultraluminous X-ray sources and SS 433 in different ranges, it was concluded that they represent a single class of objects with supercritical accretion disks (Fabrika et al., 2015). At the same time, the existing differences in the equivalent widths of the observed emission lines can be attributed to higher gas temperatures in ULX winds at lower outflow rates than in SS 433, which is at least partially confirmed by the first results of modeling the optical spectra of ULXs (Kostenkov et al., 2020).

Modeling ULX optical spectra is complex task and requires taking into account many factors. However, the similarity between ULX outflows and stellar winds allows us to use models of extended atmospheres, which are successfully used to model the spectra and determine the parameters of stellar winds. The most advanced programs developed for such modeling are the codes CMFGEN (Hillier et al., 1998) and PoWR (Hamann et al., 2006). In Kostenkov et al. (2020), we estimated temperatures and outflow rates for Holmberg II X-1, NGC 5204 X-1, NGC 4559 X-7 and UGC 6456 ULX using the CMFGEN code. The estimates of the photospheric temperature of all four objects are found to be in the narrow range $T_{ph} = 33000$–$36000$ K, and the outflow rates—$\dot{M} = 1.1$–$7.6 \times 10^{-5}\ M_\odot\ \mathrm{yr}^{-1}$. For UGC 6456 ULX, a detailed simulation of the optical spectrum was performed, which gave slightly lower values compared to those obtained from the model grids: $T_{ph} = 31250$ K vs. $T_{ph} = 33000^{+2100}_{-750}$ K and $\dot{M} = 2.7 \times 10^{-5}\ M_\odot\ \mathrm{yr}^{-1}$ vs. $\dot{M} = 7.6^{+2.0}_{-3.3} \times 10^{-5}\ M_\odot\ \mathrm{yr}^{-1}$. Note that the radiative transfer equation in these models is solved for the spherically symmetric case. In ULX winds, the spherical symmetry is broken by the presence of a funnel, and how much this affects the wind parameters will be determined in future works.

Estimates of the temperature of the SS 433's photosphere differ greatly among different authors, reaching up to 70000 K (Dolan et al., 1997), but the most plausible value is about $T_{ph} \approx 30000$ K (see the work Fabrika 2004 and references therein). Moreover, quite strong changes in the temperature of the photosphere were found depending on the precession phase: from 21000 to 45000 K (Wagner, 1986), which may be a consequence of the presence of a funnel in the wind. Temperature measurements were carried out by approximation of photometric data by the blackbody law. The matter outflow rate of SS 433 is $\dot{M}_W \sim 10^{-4}\ M_\odot\ \mathrm{yr}^{-1}$ (Fabrika, 2004). Thus, the temperature of the wind photosphere of SS 433 (despite the considerable uncertainty of its magnitude) approximately coincides with the estimates of the temperature of the ULX winds, while the outflow rates in the ULX winds are on average lower, as was assumed in the work Fabrika et al. (2015).

### 3.4. Spectroscopy of Donors of Ultraluminous X-Ray Sources

To date, the type of donor star with varying degrees of reliability has been determined in less than ten ULXs, and this was done using optical spectroscopy data.

One of the first to be classified was the companion of M 101 ULX-1 (Liu et al., 2013). The luminosity of the object most of the time is $L_X \sim 10^{37}$ erg s$^{-1}$ (Kong et al., 2004) and only in outbursts reaches $L_X \approx 3 \times 10^{39}$ erg s$^{-1}$ (Mukai et al., 2005). During outbursts, the source has a very soft X-ray spectrum, well described by a thermal component (disk) with a temperature of $90-180$ eV (Liu et al., 2013). The optical spectra of M 101 ULX-1 were obtained when the source was expected to be in the low X-ray luminosity state (although this is not confirmed), and the contribution of the accretion disk (or wind) to the optical radiation should have been minimal. The donor in the system turned out to be a Wolf–Rayet star of the WN8 subtype with a mass of $19 \pm 1$ $M_\odot$, which was determined on the basis of the wide emission lines He II 4686, 5411 Å, He I 4471, 4922, 5876, 6679 Å, N III 4634 Å observed in the spectrum and the absence of hydrogen and carbon emission lines (Liu et al. 2013). As a result of monitoring the object, the probable orbital period of the system $P = 8.2$ days was determined, lower limits on the mass of the black hole of the order of 5 $M_\odot$ were obtained, and its most probable value was given in the range of $20-30$ $M_\odot$.

In the spectrum of the ultraluminous X-ray pulsar NGC 7793 P13, the donor-owned absorption lines were found for the first time (Motch et al., 2011, 2014), which were not being found in the spectra of many other ULXs (see, for example, Cseh et al. (2013)). The absorption lines of the Balmer series (starting from Hβ), neutral helium, silicon, and other elements observed in the object's spectrum make it possible to classify the donor as a B9Ia supergiant with a mass of $18-23$ $M_\odot$ (Motch et al., 2014). Analysis of the light curve and the radial velocity curve of the emission line He II 4686 Å showed the presence of a period of $63-65$ days. Modeling of the light curve taking into account the heating of the donor star by X-ray radiation made it possible to limit the accretor mass from above to 15 $M_\odot$ (Motch et al., 2014). As discussed in previous sections, further investigations of the system led to the discovery of coherent X-ray pulsations and the identification of the compact component as a neutron star.

Absorption features detected in the spectrum of the ultraluminous X-ray pulsar NGC300 ULX-1 indicate the presence of a red supergiant (RSG) (Heida et al., 2019). This conclusion is confirmed by the authors' IR photometry in the $J$ and $H$ bands. In contrast to NGC 7793 P13, where the donor star dominates the optical spectrum of the source, in the case of NGC 300 ULX-1, the spectrum is dominated by a hot component with a temperature of at least 20 000 K and with bright emission lines, which apparently is the wind radiation from the supercritical disk or, in our opinion, less likely, the X-ray-heated part of the donor photosphere or the outer part of the accretion disk (Heida et al., 2019).

Five more ULXs are known whose donors can be red supergiants (Heida et al., 2016, 2015; Lopez et al., 2020). The authors came to this conclusion based on the data of IR-spectroscopy. However, the results of their investigation still need further confirmation, as the identification of these five ULXs was performed using only ground-based infrared images, which, although unlikely, as shown by Lopez et al. (2020), can lead to false identification in crowded fields. In total, 113 ULXs were investigated by the authors in the IR range by the photometric method, candidate counterparts were found for 38 objects, the nature of 12 sources was spectroscopically confirmed: five turned out to be nebulae, one source is not classified, one source is AGN, and five are probably ULXs with RSG-donors. From this the authors made the far-reaching conclusion that red supergiants can be donors in $4 \pm 2\%$ ULXs, which is 4 times more than predicted by evolutionary calculations (Wiktorowicz et al., 2017).

Finally, in SS 433, the companion is the A3-7 I supergiant filling its Roche lobe, which was also classified by the weak absorption spectrum observed under the bright emission spectrum of the supercritical disk (Gies et al., 2002; Hillwig et al., 2004). The contribution of the donor to the optical radiation of the object is less than 10%.

### 3.5. SS 433: Geometry of the Outflowing Gas of Supercritical Accretion Disks

As noted above, SS 433 is the only known object in our Galaxy that constantly accretes in supercritical regime. Although it is distinguished from ULXs by a very low observed luminosity in the X-ray range—of the order of $10^{36}$ erg s$^{-1}$—its real luminosity can reach $10^{40}$ erg s$^{-1}$. The reason for this may be a powerful optically thick wind outflowing from the surface of the supercritical disk (Grandi et al., 1982; Kubota et al., 2010) and blocking the radiation of the central source for the observer (who sees the object close to the plane of the accretion disk) (Fig. 4).

Many processes and phenomena that are well studied in the SS 433 system are not yet observed (or extremely rarely observed) in ultraluminous X-ray sources. Therefore, SS 433 currently provides unique opportunities for a detailed study of the physical processes occurring in supercritical accretion disks. Unfortunately, the disk itself is not observed, being covered by the photosphere of a dense wind. However, knowledge of the geometry of gas flows at different distances from the accretor (in jets, wind, extended disk) can provide important information for understanding physical processes occuring there. From X-ray, UV and optical observations, a model emerges in which the inner cavity of the funnel is surrounded by cocoons of hot gas, which re-emit quanta of the inner regions. A very large amount of data is received in the optical range. Comparable opportunities for studying the manifestations of supercritical accretion in ultraluminous X-ray sources will be achieved only after the introduction of a new generation of optical telescopes, such as the 6-m James Webb Space Telescope (JWST) or the 39-m Extremely Large Telescope (ELT) in Chile.

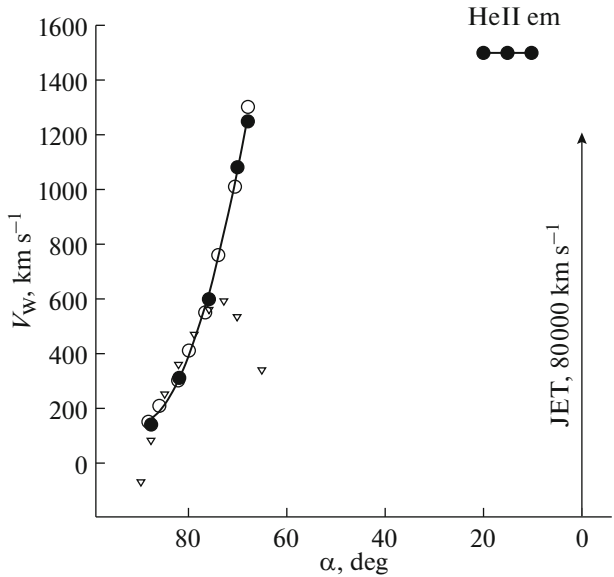

**Fig. 20.** The accretion disk wind speed, measured by the absorption components, as a function of the polar angle at different orientations of the disk. Open and filled circles on the left side are the absorption lines Hβ and He I 5015 Å. The triangles show the wind out flow in the absorption lines Fe II 5169 Å. The reverse behavior of the velocity measured by the iron lines shows that the fast wind catches up with the slow wind at long distances from SS 433. The final average wind speed along the line of sight is about $V_W \approx 340$ km s$^{-1}$. Data on the emission He II (He II cocoon) are model-dependent.

The most surprising phenomenon in SS 433 is its jets. Among the bona fide ULXs, episodic jet emissions are likely to be the energy source for the observed triple-lobed radio nebula around Holmberg II X-1 (Cseh et al., 2014, 2015). However, it is possible that radio nebulae around other ULXs are formed by the same transient jets. On the other hand, radiative hydrodynamic calculations of structure of the outflow from the surface of a supercritical disk show that the formation of collimated jets is not a mandatory phenomenon in supercritical disks: often hot gas flows with subrelativistic velocities in a fairly wide range of angles in the funnel of a denser and slower wind (see, for example, Kawashima et al. (2012)). It is likely that such a high-speed gas is just observed in many ULXs in the form of an ultra-fast outflow (Fig. 4).

In SS 433, depending on the distance from the source, the temperature of the jets, the radiation mechanism and, accordingly, the observation methods, one distinguishes between X-ray jets (about $10^{11-13}$ cm), optical jets (about $10^{14-15}$ cm), radio jets (more than $10^{15}$ cm). Extended X-ray jets are also observed at distances over $10^{17}$ cm.

An observational manifestation of jets in optical spectra is the "moving" emission lines of hydrogen and He I. The velocity in the jets measured from the lines is $0.26c$. The lines move due to changes in the inclination of the jets to the line of sight due to precession. An unexpected discovery was the detection of moving jet lines in the optical spectra of an ultraluminous super-soft X-ray source in the galaxy M 81 (Liu et al., 2015). Although the object is called ultraluminous, the (bolometric) luminosity of M81 ULS-1 in the high flux states only slightly exceeds $10^{39}$ erg s$^{-1}$, and almost all radiation is released at energies below 1 keV. The spectra of M 81 ULS-1 show two components of the Hα line that are strongly shifted relative to the also observed stationary Hα 6563 Å line. Several observations made by the authors in the blue range made it possible to measure changes in the position of the the blue-side line, which is consistent with the predicted gas velocity $0.14c-0.17c$. This indicates that the line originates in a relativistic baryon jet.

SS 433 jets are very narrow, their opening at the place where they cools down to optical temperatures and the hydrogen lines radiate (the distance corresponds to 1–3 days of flight) is $1.0°-1.5°$ (Borisov and Fabrika, 1987). X-ray jets are very short, only a few hundred seconds of flight; they show lines of highly ionized heavy elements (Marshall et al., 2002) (Fig. 7). The opening of X-ray jets is approximately $1°-2°$. The gas of the jets flies along ballistic trajectories, and the matter outflow rate in the jets is about $\dot{M}_j \approx 5\times10^{-7} \ M_\odot$ yr$^{-1}$. The kinetic luminosity of the jets is estimated to be $L_k \sim 10^{39}$ erg s$^{-1}$.

Information about the wind structure of the supercritical disk of SS 433 was obtained from the spectral and photometric data (Fabrika, 1997, 2004). The accretion disk precession allows measuring wind speed from the absorption lines as a function of the polar angle $\alpha$ measured from the disk axis. According to the kinematic model, we can study the wind only in the range of polar angles $60° < \alpha < 90°$. When the disk is oriented in such a way that the observer sees it edge-on ($\alpha = 90°$), a dense and slow wind ($V_W \approx 100$ km s$^{-1}$) is observed. As the angular distance from the disk plane increases, the wind accelerates sharply and reaches speeds of $V_W \approx 1300$ km s$^{-1}$ (Fig. 20). The measurements were carried out using the absorption components of lines with the P Cyg profile during many precession cycles. While the lines of hydrogen and He I show the same dependence on the polar angle, the line of iron Fe II 5169 Å follows this dependence only up to about 600 km s$^{-1}$, after which its radial velocity again begins to decrease and reaches the value of $V_W \approx 340$ km s$^{-1}$ at $\alpha = 60°$. The wind speed from the He II line (Fig. 20) in the region of angles $\alpha \sim 10°-20°$ was determined under the assumption that the two-peak profile of He II is formed in cocoons enveloping the base of the jets. Unlike the results obtained from Hβ, He I and Fe II, the wind speed in the line He II is not the result of direct measurements.

In addition to the well-known precession and orbital variabilities, a cycle equal to 1/7 of the precession period $P_7 = 23^d.228 \pm 0.005$ associated with a spiral shock wave was found. It is believed that this can explain the problem of viscosity in the accretion disk (Fabrika, 1997). During the time of passage of the matter through the disk, the matter inside the disk makes several revolutions, after which it freely falls on the compact object. The value of the mass function is found, which shows that the optical star is massive, and the mass of the relativistic star is estimated as $M_X \geqslant 6 M_\odot$

The ratio of the mass of the donor to the mass of the accretion disk of SS 433 is such that it leads to a significant outflow of matter through the Lagrange point L2, as a result of which an extended "circumboundary" disk is formed around the binary system, which is illuminated due to the precession of the accretion disk around the compact object (Fabrika, 1993). Using various methods of studying the envelope, it is possible to obtain estimates of the mass ratio of the accretor and donor $q = M_X/M_d$. In the recent paper Bowler (2018), the profile of stationary emission lines He I and its changes depending on the orbital phase were modeled, based on different ratios of the size of the orbit of a compact object and the size of the inner orbit of matter in an extended disk. Together with the known gas velocity in the extended disk $V = 240 \pm 10$ km s$^{-1}$ measured from the same lines He I, this allowed to obtain the value $q = 0.72 \pm 0.05$ and estimate the masses of the donor and accretor: about 21 $M_\odot$ and 15 $M_\odot$, respectively. Cherepashchuk et al. (2019), based on the constancy of the orbital period over the past 30 years, despite a significant outflow of mass and angular momentum through the L2 point (measured by the authors from VLTI+GRAVITY data), obtained a mass ratio of $q \gtrsim 0.6$. The estimates of the donor mass range from 8 $M_\odot$ to 15 $M_\odot$, and the black hole mass is $M_X \sim 5-9 M_\odot$.

The structure of the extended disk of SS 433 was studied in Waisberg et al. (2019) using data obtained with the VLTI+GRAVITY optical interferometer. Observations in the stationary line Brγ in the near-infrared region revealed an elongated structure with a size of about 1 mas (corresponding to a linear size of 5 AU). It is perpendicular to the jets and has a high rotation speed. According to the precession model of the slaved disk, the rotation occurs in the direction opposite to the precession of the jet. The authors interpreted the detected structure as an ejection in the extended disk due to centrifugal force, which implies a high efficiency of angular momentum transfer to the disk from the binary system (Waisberg et al., 2019). Note that the profile of the Brγ line has a two-peak structure, the difference between the radial velocities of the peaks reaches $\Delta V \approx 500$ km s$^{-1}$ (Fig. 21). In addition to the equatorial structure, a very extended (about 6 mas, or about 30 AU) component of the spherical wind was revealed: the binary system is completely immersed in the spherical shell optically thin in the Brγ line.

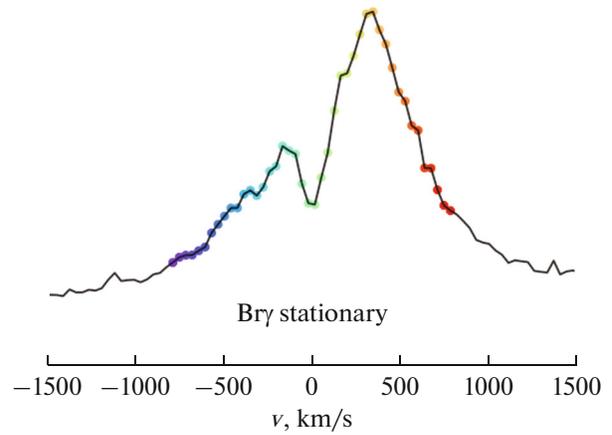

**Fig. 21.** Profile of the Brγ line as observed on July 17, 2016. With bipolar outflow, the specific angular momentum is transmitted to the extended disk. The total mass of the binary system is $M = 40 \ M_\odot$. The dots show different velocity channels (for more information, see Waisberg et al. (2019)).

## 4. CONCLUSIONS

Despite significant advances in the study of ultraluminous X-ray sources, the puzzle of the nature of these objects is still far from being solved. Now there is no doubt that accretion in most of them occurs on stellar-mass black holes or neutron stars in a supercritical mode, like SS 433. Only a relatively small part of ULXs, as well as hyperluminous X-ray sources, remain candidates for intermediate-mass black holes. Some objects, apparently, enter the supercritical mode only at the moments of outbursts of different duration.

One of the most important discoveries of the last six years was the detection of neutron stars in some ULXs by coherent pulsations of their X-ray radiation. However, the proportion of such systems among ULXs remains unclear, since pulsations are not constant and are observed not in all sources where the accretor is a neutron star. It is possible that the study of ULX donors and environments can bring us closer to the answer to this and many other questions, since evolutionary calculations predict different ages of systems and companion masses for black holes and neutron stars.

ULXs may represent a short-term but widespread stage in the evolution of a wide class of X-ray binaries. This stage can be a key stage in the formation of unique objects, such as close systems with two compact objects (black hole + black hole, black hole + neutron star, neutron star + neutron star), the probable progenitors of gravitational wave phenomena (Marchant et al., 2017).

## FUNDING

The work was supported by the Russian Foundation for Basic Research under the grant "Expansion," project No. 19-12-50215, as well as grant No. 19-02-00432.

## CONFLICT OF INTERESTS

The authors declare no conflicts of interest.

## REFERENCES

1. P. Abolmasov, S. Fabrika, O. Sholukhova, and V. Afanasiev, Astrophysical Bulletin **62**, 36 (2007).
2. P. Abolmasov, S. Fabrika, O. Sholukhova, and T. Kotani, arXiv:0809.0409 (2008).
3. M. A. Abramowicz, B. Czerny, J. P. Lasota, and E. Szuszkiewicz, Astrophys. J. **332**, 646 (1988).
4. V. K. Agrawal and A. Nandi, Monthly Notices Royal Astron. Soc. **446**, 3926 (2015).
5. J. Aird, K. Nandra, E. S. Laird, et al., Monthly Notices Royal Astron. Soc. **401**, 2531 (2010).
6. K. E. Atapin and S. N. Fabrika, Astronomy Letters **42**, 517 (2016).
7. K. Atapin and S. Fabrika, ASP Conf. Ser. **510**, 478 (2017).
8. K. Atapin, S. Fabrika, A. Medvedev, and A. Vinokurov, Monthly Notices Royal Astron. Soc. **446**, 893 (2015).
9. K. Atapin, S. Fabrika, and M. D. Caballero-García, Monthly Notices Royal Astron. Soc. **486**, 2766 (2019).
10. P. Arévalo and P. Uttley, Monthly Notices Royal Astron. Soc. **367**, 801 (2006).
11. S. Avdan, A. Vinokurov, S. Fabrika, et al., Monthly Notices Royal Astron. Soc. **455**, L91 (2016).
12. S. Avdan, A. Akyuz, A. Vinokurov, et al., Astrophys. J. **875**, 68 (2019).
13. M. Bachetti, V. Rana, D. J. Walton, et al., Astrophys. J. **778**, 163 (2013).
14. M. Bachetti, F. A. Harrison, D. J. Walton, et al., Nature **514**, 202 (2014).
15. M. Bachetti, T. J. Maccarone, M. Brightman, et al., Astrophys. J. **891**, 44 (2020).
16. M. M. Basko and R. A. Sunyaev, Astron. and Astrophys. **42**, 311 (1975).
17. M. M. Basko and R. A. Sunyaev, Monthly Notices Royal Astron. Soc. **175**, 395 (1976).
18. K. Belczynski and J. Ziolkowski, Astrophys. J. **707**, 870 (2009).
19. T. M. Belloni, arXiv:1803.03641 (2018).
20. N. V. Borisov and S. N. Fabrika, Sov. Astron. Lett. **13**, 200 (1987).
21. J. N. Bregman, J. N. Felberg, P. J. Seitzer, et al., arXiv:1205.0424 (2012).
22. M. Brightman, F. A. Harrison, F. Fürst, et al., Nature Astronomy **2**, 312 2018).
23. M. Brightman, F. A. Harrison, M. Bachetti, et al., Astrophys. J. **873**, 115 (2019).
24. W. Brinkmann, T. Kotani, and N. Kawai, Astron. and Astrophys. **431**, 575 (2005).
25. M. Brorby, P. Kaaret, and H. Feng, Monthly Notices Royal Astron. Soc. **448**, 3374 (2015).
26. M. G. Bowler, Astron. and Astrophys. **619**, L4 (2018).
27. M. D. Caballero-García, T. M. Belloni, and A. Wolter, Monthly Notices Royal Astron. Soc. **435**, 2665 (2013a).
28. M. D. Caballero-García, T. Belloni, and L. Zampieri, Monthly Notices Royal Astron. Soc. **436**, 3262 (2013b).
29. S. Carpano, F. Haberl, C. Maitra, and G. Vasilopoulos, Monthly Notices Royal Astron. Soc. **476**, L45 (2018).
30. J. Casares, P. A. Charles, D. H. P. Jones, et al., Monthly Notices Royal Astron. Soc. **250**, 712 (1991).
31. A. D. Chandra, J. Roy, P. C. Agrawal, and M. Choudhury, Monthly Notices Royal Astron. Soc. **495**, 2664 (2020).
32. A. Chashkina, P. Abolmasov, and J. Poutanen, Monthly Notices Royal Astron. Soc. **470**, 2799 (2017).
33. A. Chashkina, G. Lipunova, P. Abolmasov, and J. Poutanen, Astron. and Astrophys. **626**, A18 (2019).
34. A. M. Cherepashchuk, K. A. Postnov, and A. A. Belinski, Monthly Notices Royal Astron. Soc. **485**, 2638 (2019).
35. E. J. M. Colbert and R. F. Mushotzky, Astrophys. J. **519**, 89 (1999).
36. D. Cseh, S. Corbel, P. Kaaret, et al., Astrophys. J. **749**, id. 17 (2012).
37. D. Cseh, F. Grisé, S. Corbel, and P. Kaaret, Astrophys. J. **728**, L5 (2011).
38. D. Cseh, F. Grisé, P. Kaaret, et al., Monthly Notices Royal Astron. Soc. **435**, 2896 (2013).
39. D. Cseh, P. Kaaret, S. Corbel, et al., Monthly Notices Royal Astron. Soc. **439**, L1 (2014).
40. D. Cseh, J. C. A. Miller-Jones, P. G. Jonker, et al., Monthly Notices Royal Astron. Soc. **452**, 24 (2015).
41. T. Dauser, M. Middleton, and J. Wilms, Monthly Notices Royal Astron. Soc. **466**, 2236 (2017).
42. S. W. Davis, R. Narayan, Y. Zhu, et al., Astrophys. J. **734**, 111 (2011).
43. R. Della Ceca, F. J. Carrera, A. Caccianiga, et al., Monthly Notices Royal Astron. Soc. **447**, 3227 (2015).
44. B. De Marco, G. Ponti, G. Miniutti, et al., Monthly Notices Royal Astron. Soc. **436**, 3782 (2013).
45. J. F. Dolan, P. T. Boyd, S. Fabrika, et al., Astron. and Astrophys. **327**, 648 (1997).
46. V. Doroshenko, A. Santangelo, and L. Ducci, Astron. and Astrophys. **579**, A22 (2015).
47. V. Doroshenko, S. Tsygankov and A. Santangelo, Astron. and Astrophys. **613**, 19 (2018).

48. V. Doroshenko, S. N. Zhang, A. Santangelo, et al., Monthly Notices Royal Astron. Soc. **491**, 1857 (2020).

49. H. M. Earnshaw and T. P. Roberts, Monthly Notices Royal Astron. Soc. **467**, 2690 (2017).

50. H. P. Earnshaw, M. Heida, M. Brightman, et al., Astrophys. J. **891**, 153 (2020).

51. H. P. Earnshaw, T. P. Roberts, M. J. Middleton, et al., Monthly Notices Royal Astron. Soc. **483**, 5554 (2019).

52. H. P. Earnshaw, T. P. Roberts, and R. Sathyaprakash, Monthly Notices Royal Astron. Soc. **476**, 4272 (2018).

53. G. Fabbiano, Annual Rev. Astron. Astrophys. **27**, 87 (1989).

54. S. Fabrika, Monthly Notices Royal Astron. Soc. **261**, 241 (1993).

55. S. Fabrika, Astrophys. and Space Sci. **252**, 439 (1997).

56. S. Fabrika, Astrophysics and Space Physics Reviews **12**, 1 (2004).

57. S. N. Fabrika, in *Proc. Conf. on Accretion Processes in Cosmic Sources, St. Petersburg, Russia, 2016*, id. 46.

58. S. Fabrika and A. Mescheryakov, IAU Symp. **205**, 268 (2001).

59. S. N. Fabrika, P. K. Abolmasov, and S. Karpov, IAU Symp. **238**, 225 (2007).

60. S. Fabrika, S. Karpov, P. Abolmasov, and O. Sholukhova, IAU Symp. **230**, 278 (2006).

61. S. Fabrika, Y. Ueda, A. Vinokurov, et al., Nature Physics **11**, 551 (2015)

62. S. Fabrika, A. Vinokurov, and K. Atapin, in *Proc. The Fourteenth Marcel Grossmann Meet. On Recent Developments in Theoretical and Experimental General Relativity, Astrophysics, and Relativistic Field Theories*, Ed. by M. Bianchi, R. T. Jansen, and R. Ruffini (World Scientific Publishing Co. Pte. Ltd., 2018), 1023.

63. S. A. Farrell, N. A. Webb, D. Barret, et al., Nature **460**, 73 (2009).

64. H. Feng and P. Kaaret, Astrophys. J. **653**, 536 (2006).

65. H. Feng and P. Kaaret, Astrophys. J. **675**, 1067 (2008).

66. H. Feng, L. Tao, P. Kaaret, and F. Grisé, Astrophys. J. **831**, 117 (2016).

67. E. Filippova, M. Revnivtsev, S. Fabrika, et al., Astron. and Astrophys. **460**, 125 (2006).

68. J. Frank, A. King, and D. J. Raine, *Accretion Power in Astrophysics*, 3rd ed. (Cambridge Univ. Press, Cambridge, 2002).

69. F. Fürst, D. J. Walton, F. A. Harrison, et al., Astrophys. J. **831**, L14 (2016).

70. F. Fürst, D. J. Walton, D. Stern, et al., Astrophys. J. **834**, 77 (2017).

71. D. R. Gies, W. Huang, and M. V. McSwain, Astrophys. J. **578**, L67 (2002).

72. J. C. Gladstone, C. Copperwheat, C. O. Heinke, et al., Astrophys. J. Suppl. **206**, 14 (2013).

73. J. C. Gladstone, T. P. Roberts, and C. Done, Monthly Notices Royal Astron. Soc. **397**, 1836 (2009).

74. S. A. Grandi and R. P. S. Stone, Publ. Astron. Soc. Pacific **94**, 80 (1982).

75. F. Grisé, P. Kaaret, M. W. Pakull, and C. Motch, Astrophys. J. **734**, 23 (2011).

76. F. Grisé, P. Kaaret, S. Corbel, et al., Astrophys. J. **745**, 123 (2012).

77. F. Grisé, P. Kaaret, S. Corbel, et al., Monthly Notices Royal Astron. Soc. **433**, 1023 (2013).

78. F. Grisé, M. W. Pakull, R. Soria, et al., Astron. and Astrophys. **486**, 151 (2008).

79. F. Grisé, M. W. Pakull, R. Soria, and C. Motch, AIP Conf. Proc. **1126**, pp. 201−203 (2009).

80. E. Gotthelf, J. P. Halpern, J. Patterson, and R. M. Rich, Astron. J. **103**, 219 (1992).

81. W.-R. Hamann, G. Grafener, and A. Liermann, ASP Conf. Ser. **353**, 185 (2006).

82. M. Heida, P. G. Jonker, M. A. P. Torres, et al., Monthly Notices Royal Astron. Soc., **459**, 771 (2016).

83. M. Heida, R. M. Lau, B. Davies, et al., Astrophys. J. **883**, L34 (2019).

84. M. Heida, M. A. P. Torres, P. G. Jonker, et al., Monthly Notices Royal Astron. Soc., **453**, 3510 (2015).

85. L. M. Heil, S. Vaughan, and T. P. Roberts, Monthly Notices Royal Astron. Soc. **397**, 1061 (2009).

86. L. Hernández-García, S. Vaughan, T. P. Roberts, and M. Moddleton, Monthly Notices Royal Astron. Soc. **453**, 2877 (2015).

87. D. J. Hillier and D. L. Miller, Astrophys. J. **496**, 407 (1998).

88. T. C. Hillwig, D. R. Gies, W. Huang, et al., Astrophys. J. **615**, 422 (2004).

89. R. W. Hunstead, K. Wu, and D. Campbell-Wilson, ASP Conf. Ser. **121**, 63 (1997).

90. A. P. Igoshev and S. B. Popov, Monthly Notices Royal Astron. Soc. **473**, 3204 (2018).

91. A. F. Illarionov and R. A. Sunyaev, Astron. and Astrophys. **39**, 185 (1975).

92. A. Ingram and C. Done, Monthly Notices Royal Astron. Soc. **415**, 2323 (2011).

93. A. Ingram and M. van der Klis, Monthly Notices Royal Astron. Soc. **434**, 1476 (2013).

94. G. L. Israel, A. Belfiore, L. Stella, et al., Science **355**, 817 (2017a).

95. G. L. Israel, A. Papitto, P. Esposito, et al., Monthly Notices Royal Astron. Soc. **466**, L48 (2017b).

96. M. Jaroszynski, M. A. Abramowicz, and B. Paczynski, Acta Astronomica **30**, 1 (1980).

97. V. Jithesh, C. Anjana, and R. Misra, Monthly Notices Royal Astron. Soc. **494**, 4026 (2020).

98. P. Kaaret, H. Feng, and T. P. Roberts, Annual Rev. Astron. Astrophys. **55**, 303 (2017).

99. P. Kaaret, H. Feng, D. S. Wong, et al., Astrophys. J. **714**, L167 (2010).

100. A. R. King, M. B. Davies, M. J. Ward, et al., Astrophys. J. **552**, L109 (2001).

101. T. Kawashima and K. Ohsuga, Publ. Astron. Soc. Japan **72**, 15 (2020).

102. T. Kawashima, S. Mineshige, K. Ohsuga, and T. Ogawa, Publ. Astron. Soc. Japan **68**, 83 (2016).

103. T. Kawashima, K. Ohsuga, S. Mineshige, et al., Astrophys. J. **752**, 18 (2012).

104. I. Khabibullin, P. Medvedev, and S. Sazonov, Monthly Notices Royal Astron. Soc. **455**, 1414 (2016).

105. A. R. King, Monthly Notices Royal Astron. Soc. **393**, L41 (2009).

106. A. King and J.-P. Lasota, Monthly Notices Royal Astron. Soc. **458**, L10 (2016).

107. A. King and J.-P. Lasota, Monthly Notices Royal Astron. Soc., **485**, 3588 (2019).

108. A. King and J.-P. Lasota, Monthly Notices Royal Astron. Soc. **494**, 3611 (2020).

109. A. King, J.-P. Lasota, and W. Kluzniak, Monthly Notices Royal Astron. Soc. **468**, L59 (2017).

110. D.-W. Kim and G. Fabbiano, Astrophys. J. **721**, 1523 (2010).

111. W. Kluzniak and J.-P. Lasota Monthly Notices Royal Astron. Soc. **448**, L43 (2015).

112. A. K. H. Kong, R. Di Stefano, and F. Yuan, Astrophys. J. **617**, L49 (2004).

113. P. Kosec, C. Pinto, D. J. Walton, et al., Monthly Notices Royal Astron. Soc. **479**, 3978 (2018).

114. A. Kostenkov, A. Vinokurov, Y. Solovyeva, et al., Astrophysical Bulletin **75**, 182 (2020).

115. O. Kotov, E. Churazov, and M. Gilfanov, Monthly Notices Royal Astron. Soc. **327**, 799 (2001).

116. K. Kubota, Y. Ueda, S. Fabrika, et al., Astrophys. J. **709**, 1374 (2010).

117. I. Lehmann, T. Becker, S. Fabrika, et al., Astron. and Astrophys. **431**, 847 (2005).

118. C. Leitherer, D. Schaerer, J. D. Goldader, et al., Astrophys. J. Suppl. **123**, 3 (1999).

119. L. C.-C. Lin, C.-P. Hu, A. K. H. Kong, et al., Monthly Notices Royal Astron. Soc. **454**, 1644 (2015).

120. V. M. Lipunov, Astrophys. and Space Sci. **132**, 1 (1987).

121. G. V. Lipunova, Astronomy Letters **25**, 508 (1999).

122. J. Liu, Astrophys. J. Suppl. **192**, 10 (2011).

123. J. Liu and R. Di Stefano, Astrophys. J. **674**, L73 (2008).

124. J.-F. Liu, Y. Bai, S. Wang, et al., Nature **528**, 108 (2015).

125. J.-F. Liu, J. N. Bregman, Y. Bai, et al., Nature **503**, 500 (2013).

126. J.-F. Liu, J. Bregman, J. Miller, and P. Kaaret, Astrophys. J. **661**, 165 (2007).

127. K. M. López, M. Heida, P. G. Jonker, et al., Monthly Notices Royal Astron. Soc. **497**, 917 (2020).

128. W. Luangtip, T. P. Roberts, and C. Done, Monthly Notices Royal Astron. Soc. **460**, 4417 (2016).

129. Y. E. Lyubarskii, Monthly Notices Royal Astron. Soc. **292**, 679 (1997).

130. T. J. Maccarone, A. Kundu, S. E. Zepf, and K. L. Rhode, Nature **445**, 183 (2007).

131. P. Marchant, N. Langer, P. Podsiadlowski, et al., Astron. and Astrophys. **604**, A55 (2017).

132. H. L. Marshall, C. R. Canizares, and N. S. Schulz, Astrophys. J. **564**, 941 (2002).

133. R. G. Martin, C. Nixon, P. J. Armitage, et al., Astrophys. J. **710**, L34 (2014).

134. J. E. McClintock and R. A. Remillard, in *Compact Stellar X-ray Sources*, Ed. by W. Lewin, M. van der Klis (Cambridge Univ. Press, Cambridge, 2006) pp. 157—213 (Cambridge Astrophys. Ser., No. 39).

135. A. Medvedev and S. Fabrika, Monthly Notices Royal Astron. Soc. **402**, 479 (2010).

136. A. S. Medvedev and J. Poutanen, Monthly Notices Royal Astron. Soc. **431**, 2690 (2013).

137. M. J. Middleton and A. King, Monthly Notices Royal Astron. Soc. **470**, L69 (2017).

138. M. J. Middleton, V. Brightman, F. Pintore, et al., Monthly Notices Royal Astron. Soc. **468**, 2 (2019).

139. M. J. Middleton, L. Heil, F. Pintore, et al., Monthly Notices Royal Astron. Soc. **447**, 3243 (2015a).

140. M. J. Middleton, D. J. Walton, A. Fabian, et al., Monthly Notices Royal Astron. Soc. **454**, 3134 (2015b).

141. S. Mineo, M. Gilfanov, and R. Sunyaev, Monthly Notices Royal Astron. Soc. **419**, 2095 (2012).

142. J. Mönkkönen, S. S. Tsygankov, A. A. Mushtukov, et al., Astron. and Astrophys. **626**, A106 (2019).

143. C. Motch, M. W. Pakull, F. Grise, and R. Soria, Astronomische Nachrichten **332**, 367 (2011).

144. C. Motch, M. W. Pakull, R. Soria, et al., Nature **514**, 198 (2014).

145. S. Motta, T. Muñoz-Darias, P. Casella, et al., Monthly Notices Royal Astron. Soc. **418**, 2292 (2011).

146. K. Mukai, M. Still, R. H. D. Corbet, et al., Astrophys. J. **634**, 1085 (2005).

147. A. A. Mushtukov, A. Ingram, and M. van der Klis, Monthly Notices Royal Astron. Soc. **474**, 2259 (2018a).

148. A. A. Mushtukov, A. Ingram, M. Middleton, et al., Monthly Notices Royal Astron. Soc. **484**, 687 (2019a).

149. A. A. Mushtukov, G. V. Lipunova, A. Ingram, et al., Monthly Notices Royal Astron. Soc. **486**, 4061 (2019b).

150. A. A. Mushtukov, V. F. Suleimanov, S. S. Tsygankov, and J. Poutanen, Monthly Notices Royal Astron. Soc. **454**, 2539 (2015).

151. A. A. Mushtukov, V. F. Suleimanov, S. S. Tsygankov, and A. Ingram, Monthly Notices Royal Astron. Soc. **467**, 1202 (2017).

152. A. A. Mushtukov, S. S. Tsygankov, V. F. Suleimanov, and J. Poutanen, Monthly Notices Royal Astron. Soc. **476**, 2867 (2018b).

153. T. Okuda, G. V. Lipunova, and D. Molteni, Monthly Notices Royal Astron. Soc. **398**, 1668 (2009).

154. K. Ohsuga and S. Mineshige, Astrophys. J. **736**, 2 (2011).

155. K. Ohsuga, M. Mori, T. Nakamoto, and S. Mineshige, Astrophys. J. **628**, 368 (2005).

156. M. W. Pakull and L. Mirioni, Revista Mexicana Astron. Astrofís. Conf. Ser. **15**, 197 (2003).

157. M. W. Pakull, F. Grise, and C. Motch, IAU Symp. **230**, pp. 293— 297 (2006).

158. A. A. Panferov and S. N. Fabrika, Astronomy Reports **41**, 506 (1997).

159. D. R. Pasham and T. E. Strohmayer, Astrophys. J. **753**, 139 (2012).

160. D. R. Pasham and T. E. Strohmayer, Astrophys. J. **764**, 93 (2013).

161. D. R. Pasham and T. E. Strohmayer, Astrophys. J. **771**, 101 (2013).

162. D. R. Pasham, S. B. Cenko, A. Zoghbi, et al., Astrophys. J. **811**, L11 (2015).

163. C. Pinto, W. Alston, R. Soria, et al., Monthly Notices Royal Astron. Soc. **468**, 2865 (2017a).

164. C. Pinto, A. Fabian, M. Middleton, and D. Walton, Astronomische Nachrichten **338**, 234 (2017b).

165. C. Pinto, M. Mehdipour, D. J. Walton, et al., Monthly Notices Royal Astron. Soc. **491**, 5702 (2020).

166. C. Pinto, M. J. Middleton and A. C. Fabian, Nature **533**, 64 (2016).

167. F. Pintore and L. Zampieri, Monthly Notices Royal Astron. Soc. **420**, 1107 (2012).

168. F. Pintore, L. Zampieri, L. Stella, et al., Astrophys. J. **836**, 113 (2017).

169. F. Pintore, L. Zampieri, A. Wolter, and T. Belloni, Monthly Notices Royal Astron. Soc. **439**, 3461 (2014).

170. J. Poutanen, S. Fabrika, A. F. Valeev, et al., Monthly Notices Royal Astron. Soc. **432**, 506 (2013).

171. J. Poutanen, G. Lipunova, S. Fabrika, et al., Monthly Notices Royal Astron. Soc. **377**, 1187 (2007).

172. S. F. Portegies Zwart, H. Baumgardt, P. Hut, et al., Nature **428**, 724 (2004).

173. A. Ptak, E. Colbert, R. P. van der Marel, et al., Astrophys. J. Suppl., **166**, 154 (2006).

174. Y. Qiu, R. Soria, S. Wang, et al., Astrophys. J. **877**, 57 (2019).

175. F. Rahoui, M. Coriat, and J. C. Lee, Monthly Notices Royal Astron. Soc. **442**, 1610 (2014).

176. C. J. Ramsey, R. M. Williams, R. A. Gruendl, et al., Astrophys. J. **641**, 241 (2006).

177. F. Rao, H. Feng, and P. Kaaret, Astrophys. J. **722**, 620 (2010).

178. M. Revnivtsev, R. Burenin, S. Fabrika, et al., Astron. and Astrophys. **424**, L5 (2004).

179. M. Revnivtsev, E. Churazov, K. Postnov, and S. Tsygankov, Astron. and Astrophys. **507**, 1211 (2009).

180. M. Revnivtsev, S. Fabrika, P. Abolmasov et al., Astron. and Astrophys. **447**, 545 (2006).

181. T. P. Roberts, J. C. Gladstone, A. D. Goulding, et al., Astronomische Nachrichten **332**, 398 (2011).

182. T. P. Roberts, A. J. Levan, and M. R. Goad, Monthly Notices Royal Astron. Soc. **387**, 73 (2008).

183. G. A. Rodriguez Castillo, G. L. Israel, A. Belfiore, et al., Astrophys. J. **895**, 60 (2020).

184. C. L. Sarazin, J. A. Irwin, and J. N. Bregman, Astrophys. J. **544**, L101 (2000).

185. R. Sathyaprakash, T. R. Roberts, D. J. Walton, et al., Monthly Notices Royal Astron. Soc. **488**, L35 (2019).

186. N. I. Shakura and R. A. Sunyaev, Astron. and Astrophys. **24**, 337 (1973).

187. P. R. Shapiro, M. Milgrom, and M. J. Rees, Astrophys. J. Suppl. **60**, 393 (1986).

188. M. Shidatsu, Y. Ueda, and S. Fabrika, Astrophys. J. **839**, 46 (2017).

189. O. N. Sholukhova, S. N. Fabrika, A. V. Zharova, et al., Astrophysical Bulletin **66**, 123 (2011).

190. X. Song, D. J. Walton, G. B. Lansbury, et al., Monthly Notices Royal Astron. Soc. **491**, 1260 (2020).

191. R. Soria and A. Kong, Monthly Notices Royal Astron. Soc. **456**, 1837 (2016).

192. R. Soria, M. Cropper, M. Pakull, et al., Monthly Notices Royal Astron. Soc. **356**, 12 (2005).

193. R. Soria, K. D. Kuntz, P. F. Winkler, et al., Astrophys. J. **750**, 152 (2012).

194. R. Soria, D. T. Wickramasinghe, R. W. Hunstead, and K. Wu, Astrophys. J. **495**, L95 (1998).

195. R. Soria, K. Wu, and H. M. Johnston, Monthly Notices Royal Astron. Soc. **310**, 71 (1999).

196. A.-M. Stobbart, T. P. Roberts, and J. Wilms, Monthly Notices Royal Astron. Soc. **368**, 397 (2006).

197. T. E. Strohmayer, Astrophys. J. **706**, L210 (2009).

198. T. E. Strohmayer and R. F. Mushotzky, Astrophys. J. **586**, L61 (2003).

199. T. E. Strohmayer, R. F. Mushotzky, L. Winter, et al., Astrophys. J. **660**, 580 (2007).

200. A. D. Sutton, T. P. Roberts, D. J. Walton, et al., Monthly Notices Royal Astron. Soc. **423**, 1154 (2012).

201. A. D. Sutton, T. R. Roberts and M. J. Middleton, Monthly Notices Royal Astron. Soc. **435**, 1758 (2013).

202. A. D. Sutton, D. A. Swartz, T. P. Roberts, et al., Astrophys. J. **836**, id. 48 (2017).

203. D. A. Swartz, K. K. Ghosh, A. F. Tennant, and K. Wu, Astrophys. J. Suppl. **154**, 519 (2004).

204. H. R. Takahashi, S. Mineshige, and K. Ohsuga, Astrophys. J. **853**, 45 (2018).

205. S. Takeuchi, K. Ohsuga, and S. Mineshige S., Publ. Astron. Soc. Japan **65**, 88 (2013).

206. S. Takeuchi, K. Ohsuga, and S. Mineshige S., Publ. Astron. Soc. Japan **66**, 48 (2014).

207. L. Tao, H. Feng, F. Grisé, and P. Kaaret, Astrophys. J. **737**, 81 (2011).

208. L. Tao, H. Feng, S. Zhang, et al., Astrophys. J. **873**, 19 (2019).

209. L. Titarchuk and E. Seifina, Astron. and Astrophys. **595**, 101 (2016).

210. L. Titarchuk, N. Shaposhnikov, and V. Arefiev, Astrophys. J. **660**, 556 (2007).

211. L. J. Townsend, J. A. Kennea, M. J. Coe, et al., Monthly Notices Royal Astron. Soc. **471**, 3878 (2017).

212. S. P. Trudolyubov, Monthly Notices Royal Astron. Soc. **387**, L36 (2008).

213. S. S. Tsygankov, V. Doroshenko, A. A. Lutovinov, et al., Astron. and Astrophys. **605**, A39 (2017).

214. S. S. Tsygankov, A. A. Lutovinov, V. Doroshenko, et al., Astron. and Astrophys. **593**, A16 (2016a).

215. S. S. Tsygankov, A. A. Mushtukov, V. F. Suleimanov, and J. Poutanen, Monthly Notices Royal Astron. Soc. **457**, 1101 (2016b).

216. R. Urquhart and R. Soria, Monthly Notices Royal Astron. Soc. **456**, 1859 (2016a).

217. R. Urquhart and R. Soria, Astrophys. J. **831**, 56 (2016b).

218. G. V. Ustyugova, A. V. Koldoba, M. M. Romanova, and R. V. E. Lovelace, Astrophys. J. **646**, 304 (2006).

219. L. M. van Haaften, T. J. Maccarone, K. L. Rhode, et al., Monthly Notices Royal Astron. Soc. **483**, 3566 (2019).

220. G. Vasilopoulos, S. K. Lander, F. Koliopanos, and C. D. Bailyn, Monthly Notices Royal Astron. Soc. **491**, 4949 (2020a).

221. G. Vasilopoulos, P. S. Ray, K. C. Gendreau, et al., Monthly Notices Royal Astron. Soc. **494**, 5350 (2020b).

222. S. Vaughan, R. Edelson, R. S. Warwick, and P. Uttley, Monthly Notices Royal Astron. Soc. **345**, 1271 (2003).

223. V. A. Villar, E. Berger, R. Chornock, et al., Astrophys. J. **830**, 11 (2016).

224. A. Vinokurov, K. Atapin, and Y. Solovyeva, Astrophys. J. **893**, L28 (2020).

225. A. Vinokurov, S. Fabrika, and K. Atapin, Astrophysical Bulletin **68**, 139 (2013).

226. A. Vinokurov, S. Fabrika, and K. Atapin, Astrophys. J. **854**, 176 (2018).

227. R. M. Wagner, Astrophys. J. **308**, 152 (1986).

228. I. Waisberg, J. Dexter, P.-O. Petrucci, et al., Astron. and Astrophys. **623**, id. A47 (2019).

229. N. R. Walborn and E. L. Fitzpatrick, Publ. Astron. Soc. Pacific **112**, 50 (2000).

230. D. J. Walton, F. Furst, M. Bachetti, et al., Astrophys. J. **827**, L13 (2016).

231. D. J. Walton, F. Fürst, M. Heida, et al., Monthly Notices Royal Astron. Soc. **856**, 128 (2018).

232. D. J. Walton, F. A. Harrison, B. W. Grefenstette, et al., Astrophys. J. **793**, 21 (2014).

233. D. J. Walton, T. P. Roberts, S. Mateos, and V. Heard, Monthly Notices Royal Astron. Soc. **416**, 1844 (2011).

234. S. Wang, R. Soria, R. Urquhart, and J. Liu, Monthly Notices Royal Astron. Soc. **477**, 3623 (2018).

235. S.-S. Weng and H. Feng, Astrophys. J. **853**, 115 (2018).

236. G. Wiktorowicz, M. Sobolewska, J.-P. Lasota, and K. Belczynski, Astrophys. J. **846**, 17 (2017).

237. L. Yang, H. Feng, and P. Kaaret, Astrophys. J. **733**, 118 (2011).

238. Y. Zhang, M. Ge, L. Song, et al., Astrophys. J. **879**, 61 (2019).

*Translated by T. Sokolova*